\documentclass{aa}

\usepackage{natbib}
\bibpunct{(}{)}{;}{a}{}{,} 
\usepackage{graphics}   
\usepackage{graphicx}   
\usepackage{epstopdf}
\usepackage[varg]{txfonts}
\usepackage{pdflscape}
\usepackage{hyperref}   
\usepackage[flushleft]{threeparttable}

\hypersetup{
    bookmarks=true,         
    unicode=false,          
    colorlinks=true,       
    linkcolor=blue,          
    citecolor=blue,        
    filecolor=blue,      
    urlcolor=blue           
}

\begin{document} 

   \title{Non standard s-process in massive rotating stars}
   \subtitle{Yields of $10-150$ $M_{\odot}$ models at $Z=10^{-3}$ }

   \author{
   Arthur Choplin
          \inst{1},
          Raphael Hirschi
          \inst{2,3,4},
           Georges Meynet
           \inst{1},
          Sylvia Ekstr\"{o}m
          \inst{1},
          Cristina Chiappini
          \inst{5}
          \and
          Alison Laird
          \inst{4,6}          
                    }

 \authorrunning{Choplin et al.}

        \institute{Geneva Observatory, University of Geneva, Maillettes 51, CH-1290 Sauverny, Switzerland\\
        e-mail: arthur.choplin@unige.ch
        \and Astrophysics Group, Lennard-Jones Labs 2.09, Keele University, ST5 5BG, Staffordshire, UK
        \and Kavli Institute for the Physics and Mathematics of the Universe (WPI), University of Tokyo, 5-1-5
Kashiwanoha, Kashiwa, 277-8583, Japan
        \and UK Network for Bridging the Disciplines of Galactic Chemical Evolution (BRIDGCE)
        \and Leibniz-Institut f\"{u}r Astrophysik Potsdam, An der Sternwarte 16, 14482, Potsdam, Germany
        \and Department of Physics, University of York, York YO10 5DD, United Kingdom
                }
                
   \date{Received / Accepted}

  \abstract
   {Recent studies show that rotation significantly affects the s-process in massive stars.} 
   {We provide tables of yields for non-rotating and rotating massive stars between 10 and 150 $M_{\odot}$ at $Z=10^{-3}$ ([Fe/H] $=-1.8$). Tables for different mass cuts are provided. The complete s-process is followed during the whole evolution with a network of 737 isotopes, from Hydrogen to Polonium. 
   }
   {A grid of stellar models with initial masses of 10, 15, 20, 25, 40, 60, 85, 120 and 150 $M_{\odot}$ and with an initial rotation rate of both 0 or 40\% of the critical velocity was computed. Three extra models were computed in order to investigate the effect of faster rotation (70\% of the critical velocity) and of a lower $^{17}$O($\alpha,\gamma$) reaction rate.}
   {At the considered metallicity, rotation has a strong impact on the production of s-elements for initial masses between 20 and 60 $M_{\odot}$. In this range, the first s-process peak is boosted by $2-3$ dex if rotation is included. Above 60 $M_{\odot}$, s-element yields of rotating and non-rotating models are similar. 
   Increasing the initial rotation from 40 \% to 70 \% of the critical velocity enhances the production of $40 \lesssim Z \lesssim 60$ elements by $\sim 0.5-1$ dex.
   Adopting a reasonably lower $^{17}$O($\alpha,\gamma$) rate in the fast rotating model (70 \% of the critical velocity) boosts again the yields of s-elements with $55 \lesssim Z  \lesssim 82$ by about 1 dex. In particular, a modest amount of Pb is produced. Together with s-elements, some light elements (particularly fluorine) are strongly overproduced in rotating models.
   }
   {}

   \maketitle

	\titlerunning{Spinstars and CEMP-no stars}
	\authorrunning{A. Choplin et al.}

\section{Introduction}\label{intro}

The standard view of the s-process in massive stars is that it occurs in He- and C-burning regions and contributes to the production of elements up to about $A=90$, hence giving only s-elements up to the first peak, at $N=50$, where $N$ is the number of neutrons \citep[e.g.][and references therein]{peters68, couch74,lamb77, langer89, raiteri91a, raiteri91b, raiteri93, kappeler11}. In standard models of massive stars, both the neutron source (mainly $^{22}$Ne) and the seed (mainly $^{56}$Fe) decrease with initial metallicity while the main neutron poison ($^{16}$O) remains similar whatever the metallicity, leading to a threshold of about $Z/Z_{\odot}=10^{-2}$ below which the s-process becomes negligible \citep{prantzos90}. 

\cite{meynet06} and \cite{hirschi07} suggested that this picture would be modified in rotating stars because of the rotational mixing operating between the H-shell and He-core during the core helium burning phase. The abundant $^{12}$C and $^{16}$O isotopes in the convective He-burning core are mixed to the H-shell, boosting the CNO cycle and forming primary $^{14}$N \citep[e.g.][]{meynet02a, ekstrom08}. The $^{14}$N is mixed back into the convective He-burning core and allows the synthesis of extra $^{22}$Ne, via the reaction chain $^{14}$N($\alpha,\gamma$)$^{18}$F(,$e^+ \nu_e$)$^{18}$O($\alpha,\gamma$)$^{22}$Ne. The growth of the convective He-burning core also helps reaching layers that had been previously enriched in $^{14}$N.
Neutrons are finally released by the $^{22}$Ne($\alpha,n$)$^{25}$Mg reaction. $^{22}$Ne production in rotating stars is extensively discussed in \cite{frischknecht16} \citep[section 3.1, their Fig.~2 to 5. See also Fig.~1 of][for a schematic view of this mixing process]{choplin16}.
By investigating the effect of rotation in a 25 $M_{\odot}$ model, \cite{pignatari08} have shown that rotational mixing would allow the production of s-elements up to  $A \simeq 140$. Since then, a few studies \citep[e.g.][F12 and F16 hereafter]{frischknecht12,frischknecht16} started to build a picture of the s-process in massive rotating stars, by computing models of different masses ($15<M<40$ $M_{\odot}$) and metallicities ($10^{-7}<Z<Z_{\odot}$) while following the complete s-process during the evolution. 
So far, the most complete s-process study from rotating massive star models was carried out in F16. They computed 29 non-rotating and rotating models of 15, 20, 25 and 40 $M_{\odot}$, with metallicities of $Z = 0.014$, $10^{-3}$, $10^{-5}$ and $10^{-7}$ and with a nuclear network of 613 or 737 isotopes, depending on the burning phase.

Several observational signatures tend to support the view of an \textit{enhanced s-process} in massive rotating stars. 

The first one is the globular cluster NGC 6522, located in the galactic bulge and possibly being about 12.5 Gyr old \citep{kerber18}. It contains eight stars whose pattern is enriched in s-elements \citep{barbuy09} and consistent with the yields of massive rotating models \citep{chiappini11}.

The second signature regards the iron-poor low mass stars enriched in s-elements in the halo of the Milky Way. Using an inhomogeneous galactic chemical evolution model, \cite{cescutti13} have shown that the observed scatter in the [Sr/Ba] ratio of normal (i.e. not enriched in carbon, see next discussion) halo stars with [Fe/H] $< -2.5$ can be reproduced if including yields from fast rotating massive stars. 

The third one regards some of the [Fe/H] $\lesssim -4$ stars enriched is s-elements. At such a low metallicity, asymptotic giant branch (AGB) stars might not have contributed yet to the chemical enrichment. HE 1327-2326 \citep{aoki06, frebel06b, frebel08} has [Fe/H] $=-5.7$, [Sr/Fe] $= 1.08$, [Ba/Fe] $<1.39$ and is enriched in light elements (C, N, O, Na, Mg, Al) relatively to Fe. As discussed in the Sect.~7.2 of \cite{maeder15a}, this is consistent with the ejecta of a fast rotating low-metallicity massive star, where a strong mixing between H- and He-burning zones occurred, triggering the synthesis of a variety of elements, including Sr and Ba. 

A fourth signature concerns the CEMP-s stars that are Carbon-Enhanced Metal-Poor stars enriched in s-elements \citep{beers05}. CEMP-s stars are mostly found at [Fe/H] $>-3$ \citep[e.g.][]{yong13, norris13}. Some significantly s-rich stars also exist at [Fe/H] $<-3$, like HE 1029-0546 or SDSSJ1036+1212 with [Fe/H] around $-3.3$ \citep{behara10, aoki13,hansen15}. The peculiar chemical pattern of such stars is generally considered as acquired from a AGB star companion during a mass transfer (or wind mass transfer) episode \citep{stancliffe08, lau09,bisterzo10,bisterzo12,lugaro12,abate13,abate15b,abate15a,hollek15}. A consequence of such a scenario is that CEMP-s stars should mostly be in binary systems, which seems to be the case for most CEMP-s since they show radial velocity variations \citep{lucatello05,starkenburg14,hansen16a}. 
Nevertheless, some CEMP-s stars are very likely single stars \citep[4 out of 22 in the sample of][]{hansen16a}, challenging the AGB scenario. The yields of a fast rotating 25 $M_{\odot}$ model can reproduce the pattern of 3 out of the 4 apparently single CEMP-s stars \citep{choplin17letter}. It is also not excluded that some CEMP-s stars in binary systems show the nucleosynthetic signature of massive rotating stars since massive rotating stars could have enriched the cloud in which the binary system formed. 
On the other hand, single CEMP-s stars may be explained by the AGB scenario anyway since (1) single CEMP-s stars might have lost their companion or (2) they might be in a binary system with very long period, explaining the non-detection of radial velocity variation. 

Extensive and homogeneous grids of massive stellar models including rotation and full s-process network are needed to further investigate the role of such stars in the chemical enrichment of the universe.

In this work, we study the impact of the rotation on the s-element production at a metallicity $Z=10^{-3}$ in mass fraction and in the range $10 - 150$ $M_{\odot}$. We focus on one metallicity but extend significantly the range of mass compared to the study of F16. It allows us to draw a more complete picture of the s-process in massive stars, at the considered metallicity. We investigate also the impact of a faster initial rotation and a lower $^{17}$O($\alpha,\gamma$)$^{21}$Ne reaction rate. Sect.~\ref{ingredients} describes the physical ingredients used throughout this work. Results are presented in Sect.~\ref{nuc}. In Sect.~\ref{effmcut} we investigate the effect of the mass cut and describe the table of yields. Section~\ref{concl} presents the conclusions and additional discussions.

\begin{table}
\scriptsize{
\caption{Initial mass (column 1), model label (column 2), initial ratio of surface velocity to critical velocity (column 3), time-averaged surface velocity during the MS phase (column 4), final nuclear phase computed (column 5), total lifetime (column 6) and final mass (column 7). \label{table:1}}
\begin{center}
\resizebox{8.5cm}{!} {
\begin{threeparttable}
\begin{tabular}{c|l|cr|ccc} 
\hline 
\hline 
M$_{\rm ini}$  		& Model & $\upsilon_{\rm ini}/\upsilon_{\rm crit}$ & $\langle \upsilon \rangle_{\rm MS} $ & phase & $\tau$ & M$_{\rm fin}$ \\ 
$[M_{\odot}$]    &	      &					      & [km s$^{-1}$] 	 &    & [Myr] & $[M_{\odot}$] \\
\hline 
10	& 10s0		&		0.0		&	0		 	&end C    & 23.4	& 9.9\\
10	& 10s4		&		0.4		&	214		 	&end C    & 27.1	& 9.8\\
\hline 
15	& 15s0		&		0.0		&	0		 	&end Ne  & 13.0	& 14.8	\\
15	& 15s4		&		0.4		&	234		 	&end Ne  & 15.4	& 14.3	\\
\hline 
20	& 20s0		&		0.0		&	0		 	&end Ne  & 9.32	& 19.9	\\
20	& 20s4		&		0.4		&	260		 	&end Ne  & 10.9	& 17.4	\\
\hline 
25	& 25s0		&		0.0		&	0		 	&end Ne    & 7.61	& 24.7\\
25	& 25s4		&		0.4		&	281		 	&end Ne    & 8.81	& 16.7\\
25	& 25s7		&		0.7		&	490	 		&end Ne    & 9.20	& 16.2\\
25	& 25s7B$^{a}$&		0.7		&	490		 	&end Ne    & 9.20	& 16.0\\
\hline  
40	& 40s0		&		0.0		&	0		 	&end Ne	   & 5.24	& 34.1 \\
40	& 40s4		&		0.4		&	332		 	&end Ne	   & 5.97	& 24.6 \\
 \hline
60	& 60s0		&		0.0		&	0		 	&end Ne	   & 4.11	& 44.2\\
60	& 60s4		&		0.4		&	375		 	&end Ne      & 4.62	& 40.8 \\
 \hline
85	& 85s0		&		0.0		&	0		 	&end Ne	& 3.49	& 59.3\\
85	& 85s4		&		0.4		&	403		 	&end Ne	& 3.88	& 58.3\\
\hline
120	& 120s0	&		0.0		&	0		 	&end Ne	& 3.06	& 82.4\\
120	&120s0B$^{a}$&	0.0		&	0		 	&end Ne	& 3.06	& 83.2\\
120	& 120s4	&		0.4		&	423		 	&end He	& 3.36	& 85.8\\
\hline
150	& 150s0	&		0.0		&	0		 	&end Ne	& 2.85	& 100.3\\
150	& 150s4	&		0.4		&	432		 	&end Ne	& 3.14	& 99.6\\
\hline
\end{tabular}
\begin{tablenotes}
            \item[a] Models computed with the rate of $^{17}$O($\alpha,\gamma$) divided by 10.
\end{tablenotes}
\end{threeparttable}
}
\end{center}
}
\end{table}

   \begin{figure*}
   \centering
   \begin{minipage}[c]{.49\linewidth}
       \includegraphics[scale=0.47]{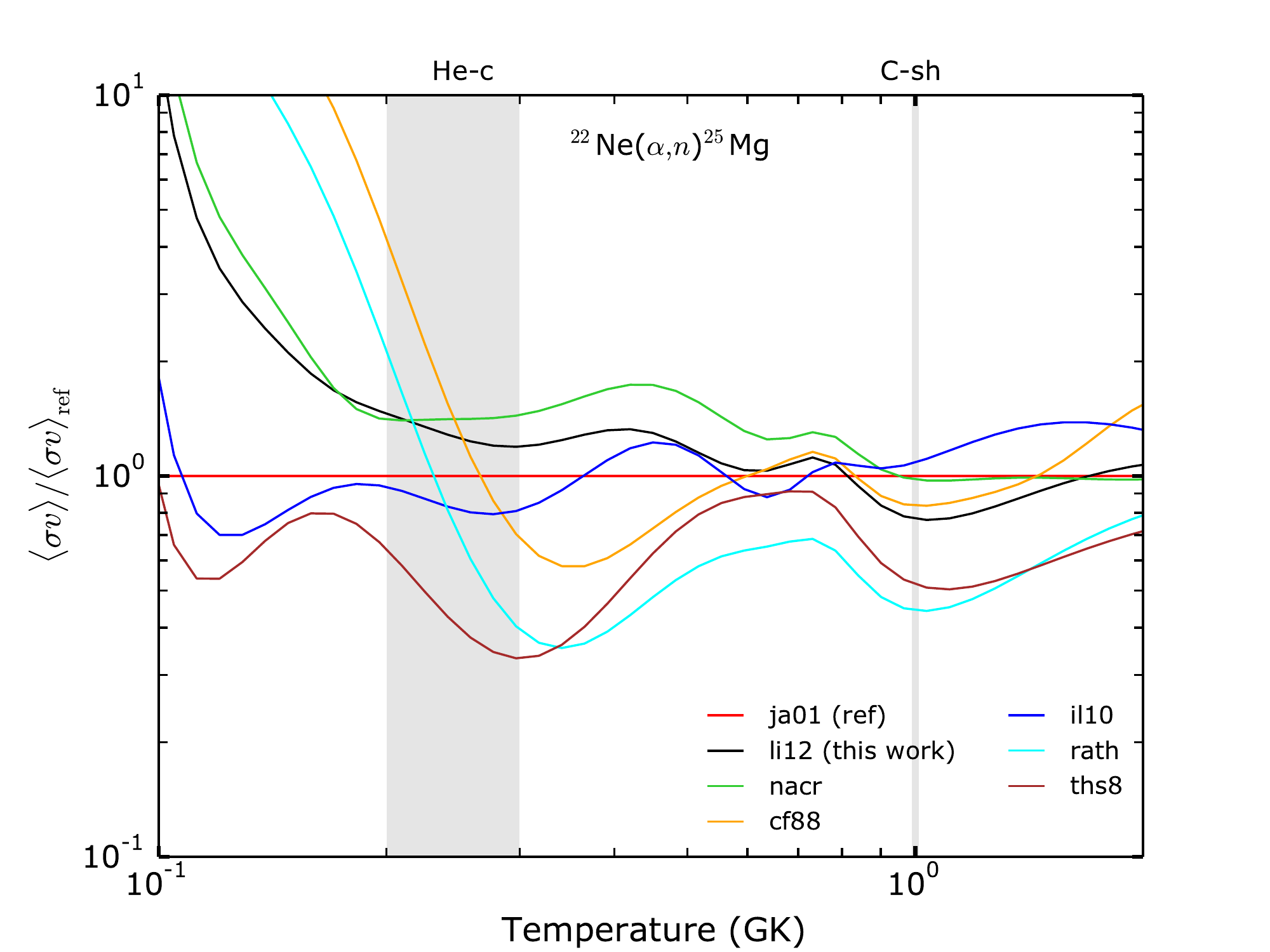}
   \end{minipage}
   \begin{minipage}[c]{.49\linewidth}
      \includegraphics[scale=0.47]{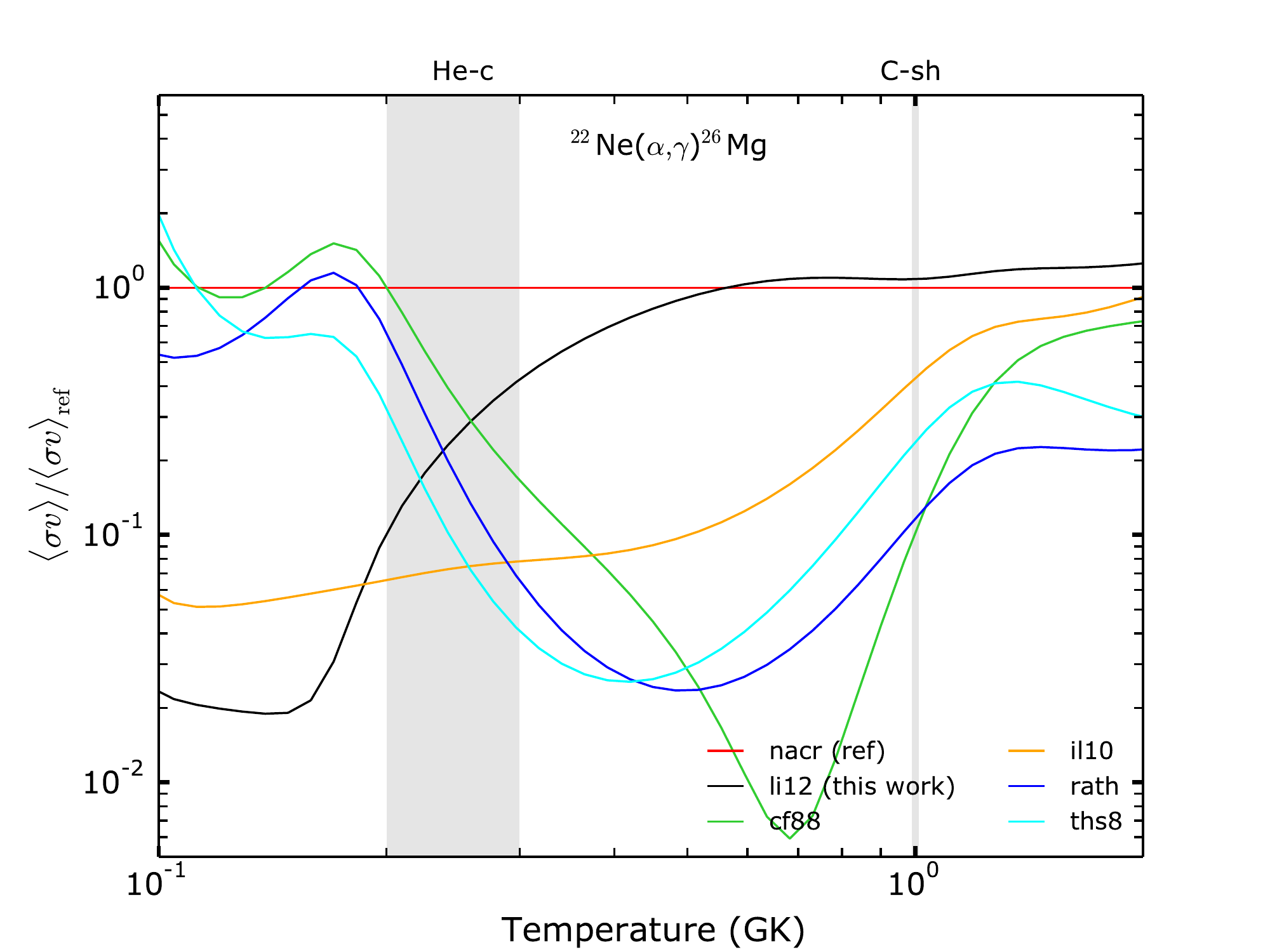}
   \end{minipage}
   \begin{minipage}[c]{.49\linewidth}
      \includegraphics[scale=0.47]{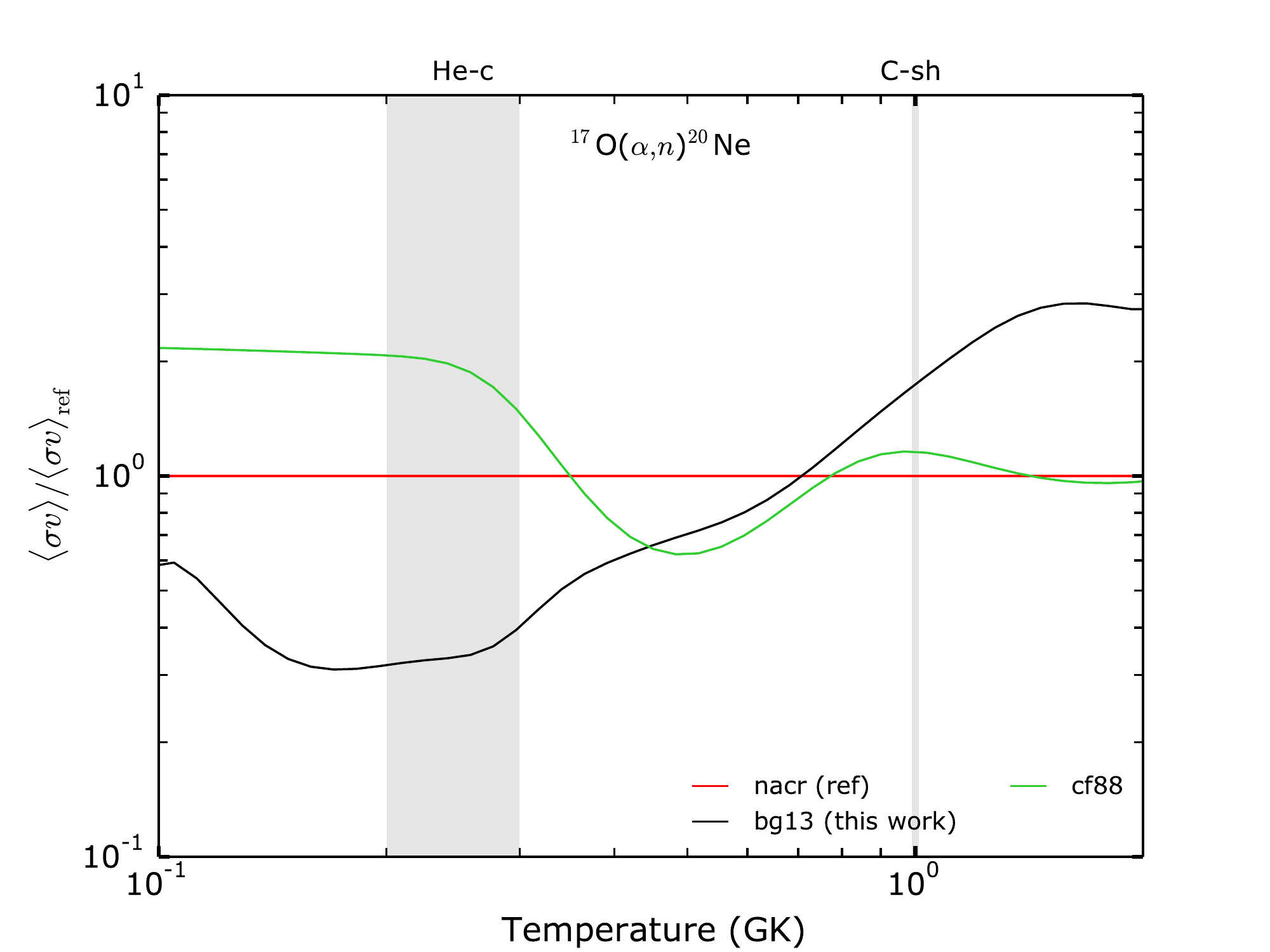}
   \end{minipage}
   \begin{minipage}[c]{.49\linewidth}
      \includegraphics[scale=0.47]{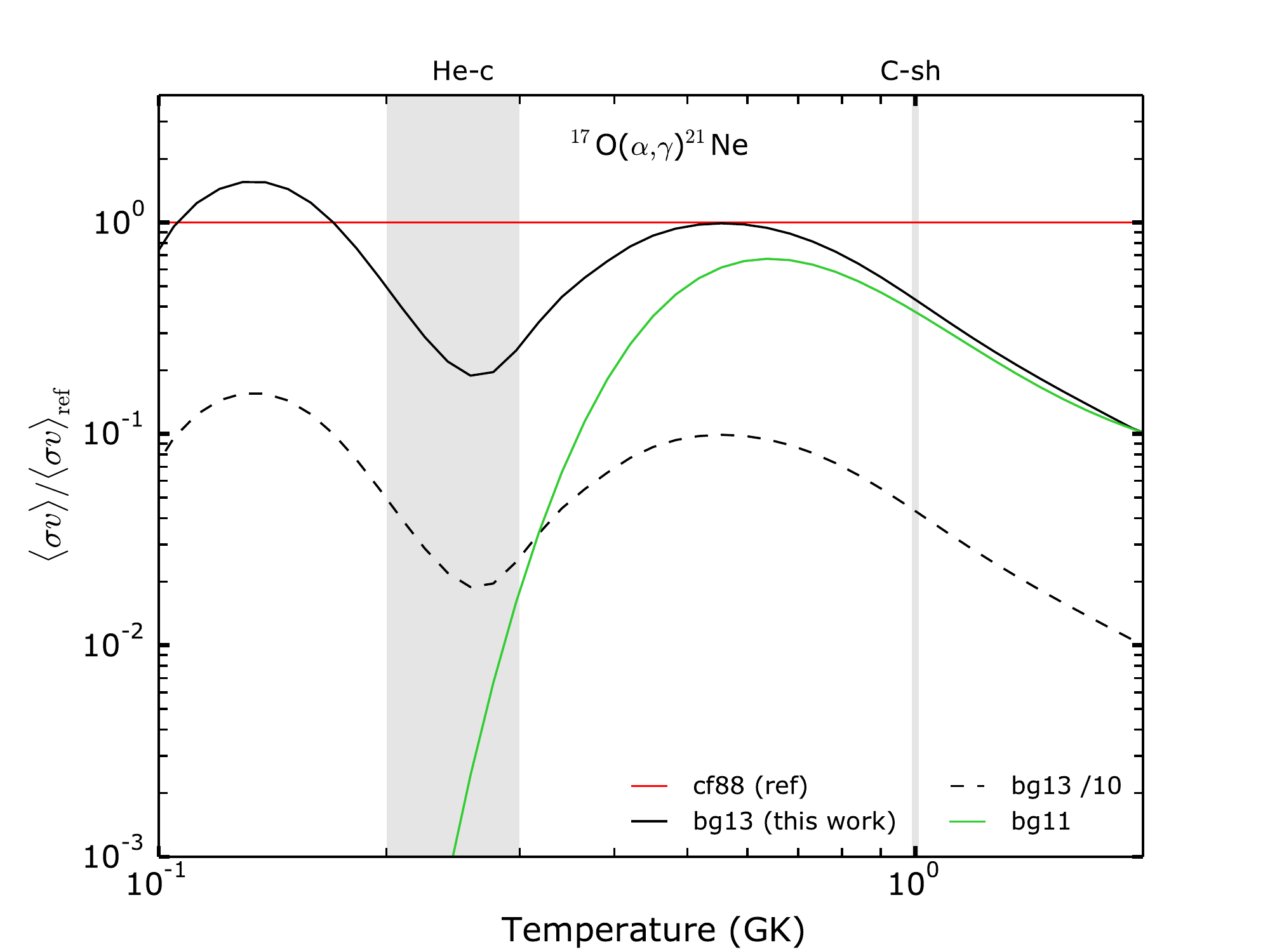}
   \end{minipage}
   \caption{Comparison between different sources of the important reaction rates for the s-process in massive stars as a function of the temperature. 'ja01': \cite{jaeger01}, 'nacr': \cite{angulo99}, 'li12': \cite{longland12}, 'cf88': \cite{caughlan88}, 'il10': \cite{iliadis10a}, 'rath': \cite{rauscher00}, 'ths8': \cite{cyburt10}, 'bg11': \cite{best11}, 'bg13': \cite{best13}. Note that the \cite{best11} rate is an experimental lower limit. 
 The shaded bands indicate the approximate ranges of temperature of interest for the s-process in massive stars: the first area named 'He-c' is associated to the s-process during the core helium burning phase and the second area ('C-sh') during carbon shell burning.}
\label{figcomprates}
    \end{figure*}

\section{Physical ingredients}\label{ingredients}

\subsection{Input parameters}\label{inputs}

We used the Geneva stellar evolution code \citep{eggenberger08}. 
The models were computed at $Z=10^{-3}$ ([Fe/H] $= -1.8$) with initial masses of 10, 15, 20, 25, 40, 60, 85, 120 and 150 $M_{\odot}$. 
The initial rotation rate on the zero-age main-sequence (ZAMS), $\upsilon_{\rm ini}/\upsilon_{\rm crit}$\footnote{$\upsilon_{\rm ini}$ is the initial equatorial velocity and $\upsilon_{\rm crit}$ is the initial equatorial velocity at which the gravitational acceleration is balanced by the centrifugal force. It is defined as $\upsilon_{\rm crit} = \sqrt{ \frac{2GM}{3R_{\rm pb}}}$ where $R_{\rm pb}$ is the the polar radius at the break-up velocity velocity \citep[see][]{maeder00a}.}
is $0$, $0.4$ or $0.7$. 
Only the 25 $M_{\odot}$ was computed with $\upsilon_{\rm ini}/\upsilon_{\rm crit} = 0.7$. As in \cite{ekstrom12} and \cite{georgy13}, we use $\upsilon_{\rm ini}/\upsilon_{\rm crit} = 0.4$ for the grid. It corresponds well to the peak of the velocity distribution of the sample of 220 young main-sequence B-type stars of \citet[their Fig.~6]{huang10}. At lower metallicities, stars are more compact and the mass loss by line driven winds is weaker so that the removal of angular momentum during evolution is smaller. Consequently, for a given $\upsilon_{\rm ini}/\upsilon_{\rm crit}$ ratio, lower metallicity stars have higher surface rotational values during the Main-Sequence phase \citep{maeder01}.

At the metallicity considered here, F16 computed (with the same stellar evolution code) non-rotating 15, 20 and 25 $M_{\odot}$ and 15, 20, 25 and 40 $M_{\odot}$ models with $\upsilon_{\rm ini}/\upsilon_{\rm crit}=0.4$. These models were computed again in the present work with the latest version of the code and with updated nuclear reaction rates (see below, the present section). A comparison of the yields is done in Sect.~\ref{rotmodels}. Our models are generally stopped at the end of the neon photo-disintegration phase. Only the 10 $M_{\odot}$ models are stopped at the end of C-burning and the rotating 120 $M_{\odot}$ is stopped at the end of He-burning.
Computing the advanced stages is important since the s-process occurs in the C-burning shell \citep[also in the He-burning shell to a smaller extent,][]{the07}. However, the contribution from He-core burning dominates both in non-rotating and rotating models  and the C-shell contribution decreases quickly with initial metallicity (F16, especially their Figure~13).
Table \ref{table:1} shows the initial properties of the models computed in this work as well as the final nuclear phase computed, the total lifetimes, and the final masses.

The nuclear network is fully coupled to the evolution and used throughout all of it. It comprises 737 isotopes, from Hydrogen to Polonium ($Z=84$). The size of the network is similar to the network used in \cite{the00}, F12, and F16, and allows to follow the complete s-process.
At the end of evolution, before computing stellar yields, unstable isotopes are decayed to stable ones.

The initial composition of metals (elements heavier than helium\footnote{The initial helium mass fraction $Y$ is calculated according to the relation $Y=Y_\text{p} + \Delta Y / \Delta Z \times Z$ where $Z$ is the metallicity, $Y_\text{p}$ the primordial helium abundance and $\Delta Y / \Delta Z = (Y_{\odot} - Y_\text{p})/Z_{\odot}$ the average slope of the helium-to-metal enrichment law.  We set $Y_\text{p}=0.248$, according to \cite{cyburt03}. We use $Z_{\odot}=0.014$ and $Y_{\odot} = 0.266$ as in \cite{ekstrom12}, derived from \cite{asplund05}. The initial mass fraction of hydrogen is deduced from $1-Y-Z = 0.752$.}) is $\alpha$-enhanced (we refer to Sect. \textsection 2.1 of F16 where more details are given).

Opacity tables are computed with the OPAL tool\footnote{\url{http://opalopacity.llnl.gov}}. At low temperature,  the opacities from \cite{ferguson05} are used to complement the OPAL tables. Radiative mass-loss rates are from \cite{vink01} when $\log T_\text{eff} \geq 3.9$ and if $M_{\rm ini} > 15$ $M_{\odot}$. Otherwise, they are from \cite{jager88}. For rotating models, the radiative mass-loss rate are corrected with the factor described in \cite{maeder00a}. Following \cite{ekstrom12}, the mass-loss rate is increased by a factor of 3 when the luminosity of any layer in the stellar envelope becomes higher than five times the Eddington luminosity.

The Schwarzschild criterion is used for convection.
During the H- and He-burning phases, overshoot is considered: the size of the convective core is extended by $d_{\rm over} = \alpha H_{\rm P}$ where $H_{\rm P}$ is the pressure scale height and $\alpha = 0.1$. $\alpha$ was calibrated so as to reproduce the observed MS width of stars with $1.35<M<9$ $M_{\odot}$ \citep{ekstrom12}.
Rotation is included according to the shellular theory of rotation \citep{zahn92}. 
The angular momentum is transported according to an advection-diffusion equation \citep{chaboyer92} which is fully solved during the Main-Sequence. Only the diffusive part of the equation is solved after the Main-Sequence.
For chemicals species, the combination of meridional circulation and horizontal turbulence can be described as a pure diffusive process \citep{chaboyer92}. The associated diffusion coefficient is
\begin{equation}
D_{\rm eff} = \frac{1}{30}\frac{\textbar r U(r) \textbar ^2}{D_{\rm h}}
\label{deff}
\end{equation}
with $U(r)$ the amplitude of the radial component of the meridional velocity \citep{maeder98} and $D_{\rm h}$ the horizontal shear diffusion coefficient from \cite{zahn92}.
The equation for the transport of chemical elements is therefore purely diffusive with a total diffusion coefficient $D_{\rm tot} = D + D_{\rm eff}$ where $D$ is the sum of the various instabilities (convection, shear...). After the Main-Sequence, the advective effects are not considered so that $D_{\rm eff} = 0$.
The secular shear diffusion coefficient is from \cite{talon97}. It is expressed as

\begin{equation}
D_{\rm shear} = f_{\rm energ} \frac{ H_{\rm p}}{g\delta} \frac{ K +D_{\rm h}}{(\nabla_{\rm ad} - \nabla_{\rm rad}) + \frac{\varphi}{\delta}\nabla_\mu(\frac{K}{D_{\rm h}}+1) } \left(\frac{9 \pi}{32} \Omega \frac{\text{d} \ln \Omega}{\text{d} \ln r}\right)^2.
\label{dshtz97}
\end{equation}
The efficiency of the shear is calibrated with the $f_{\rm energ}$ parameter. We set $f_{\rm energ} = 4$, which is the value needed for a 15 $M_{\odot}$ model at solar metallicity and with $\upsilon_{\rm ini} = 300$ km~s$^{-1}$ to obtain an enhancement of the surface N abundance by a factor of 3 at core H depletion \citep[a similar calibration is done in e.g.][]{heger00, chieffi13}. Such an surface enrichment agrees qualitatively with observation of $10 - 20$ $M_{\odot}$ rotating stars \citep[e.g.][]{gies92, villamariz05, hunter09}.

Except for some nuclear rates, we used the same inputs as those used in F16 so that the interested reader can refer to this work for further details. 
Table~\ref{table:2} lists the rates important for the s-process that were updated in the present work.

In the stellar evolution code, the rates in their analytical form \citep{rauscher00} are used. The new rates of $^{17}$O($\alpha,\gamma$) and  $^{17}$O($\alpha,n$) from \cite{best13}, used in stellar evolution models for the first time, are only tabulated. As a consequence, we derived the analytical form of these rates\footnote{more details here: http://nucastro.org/forum/viewtopic.php?id=22}. We checked that the difference between the fit and the tabulated rate was less than $5\%$.
The rates of $^{17}$O($\alpha,\gamma$), $^{17}$O($\alpha,n$), $^{22}$Ne($\alpha,\gamma$) and  $^{22}$Ne($\alpha,n$) are still uncertain in the range of temperature of interest for the s-process in massive stars \citep[e.g][]{best11, nishimura14}. Fig.~\ref{figcomprates} compares the different available rates in the literature for these four reactions. In the range of temperature of interest for us (mainly $0.2 - 0.3$ GK, corresponding to the temperature of the helium burning core), the most uncertain rate is $^{17}$O($\alpha,\gamma$). It varies by about 3 dex from the rate of \cite{caughlan88} to the rate of \cite{best11} (see the bottom right panel of Fig.~\ref{figcomprates}). 
This motivated us to test the impact of a lower $^{17}$O($\alpha,\gamma$) rate in some models. We tried a rate divided by 10 (dotted line in Fig.~\ref{figcomprates}) for the fast rotating 25 $M_{\odot}$ and non-rotating 120 $M_{\odot}$ models.

\begin{table}[h]
\scriptsize{
\caption{List of the updated reactions important for the s-process. Rates used in F12 and F16 (column 1), rates used in the present work (column 2). \label{table:2}}
\begin{center}
\resizebox{8.5cm}{!} {
\begin{tabular}{c|cc} 
\hline 
\hline 
Reaction  		& F12, F16 & This work \\ 
\hline 
 $^{12}$C($\alpha,\gamma$)$^{16}$O	&	\cite{kunz02}   &   \cite{xu13}			\\
 $^{13}$C($\alpha,n$)$^{16}$O			&	\cite{angulo99}   &    \cite{guo12}		\\
 $^{14}$N($\alpha,\gamma$)$^{18}$F	&	\cite{angulo99}   &    \cite{iliadis10a}		\\
 $^{18}$O($\alpha,\gamma$)$^{22}$Ne	&	\cite{angulo99}   &    \cite{iliadis10a}		\\
 $^{17}$O($\alpha,\gamma$)$^{21}$Ne	&	\cite{caughlan88}   &    \cite{best13}		\\
 $^{17}$O($\alpha,n$)$^{20}$Ne		&	\cite{angulo99}   &    \cite{best13}		\\
 $^{22}$Ne($\alpha,\gamma$)$^{26}$Mg	&	\cite{angulo99}   &    \cite{longland12}	\\
 $^{22}$Ne($\alpha,n$)$^{25}$Mg		&	\cite{jaeger01}   &    \cite{longland12}	\\
\hline 

\end{tabular}
}
\end{center}
}
\end{table}

\subsection{Yields and production factors}\label{secyields}

The yields provided contain a contribution from the wind and a contribution from the supernova. The yields from the supernova depends on the mass cut\footnote{At the time of the supernova, the mass cut delimits the part of the star which is expelled from the part which is locked into the remnant. The mass cut is equal to the mass of the remnant.} $M_{\rm cut}$. 
Explosive nucleosynthesis, which is not considered here, will mostly affect the iron-group elements in the innermost layers of the star \citep{woosley95, thielemann96, limongi03b, nomoto06, heger10} and is not expected to strongly modify the yields of s-elements \citep{rauscher02, tur09}. Our results hence provide good predictions for the yields of light nuclei and s-process nuclei.
The yield of an isotope $i$ is calculated according to the relation

   \begin{figure*}
   \centering
   \begin{minipage}[c]{.49\linewidth}
   \centering
      \includegraphics[scale=0.45]{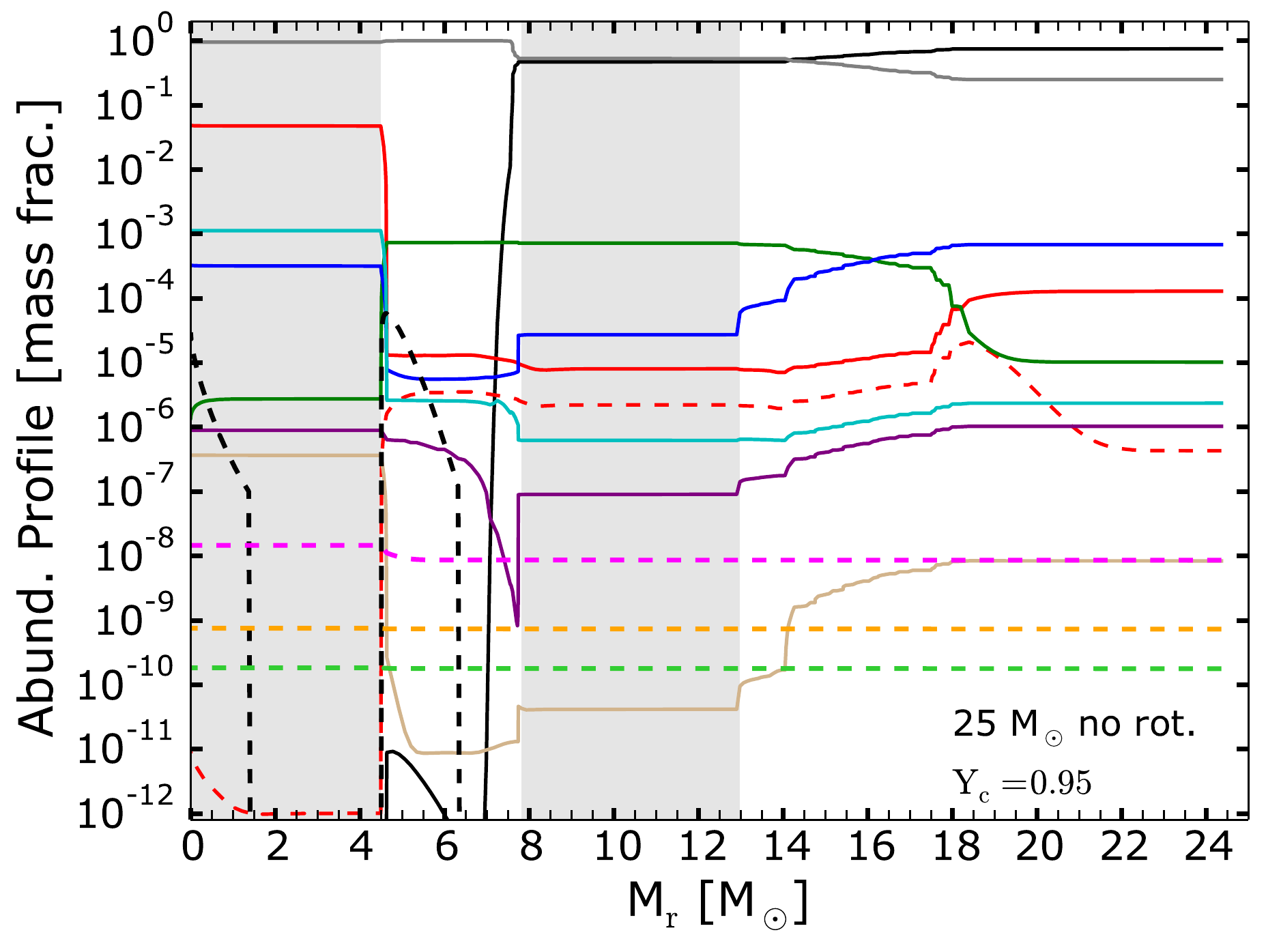}
   \centering
   \end{minipage}
   \begin{minipage}[c]{.49\linewidth}
   \centering
      \includegraphics[scale=0.45]{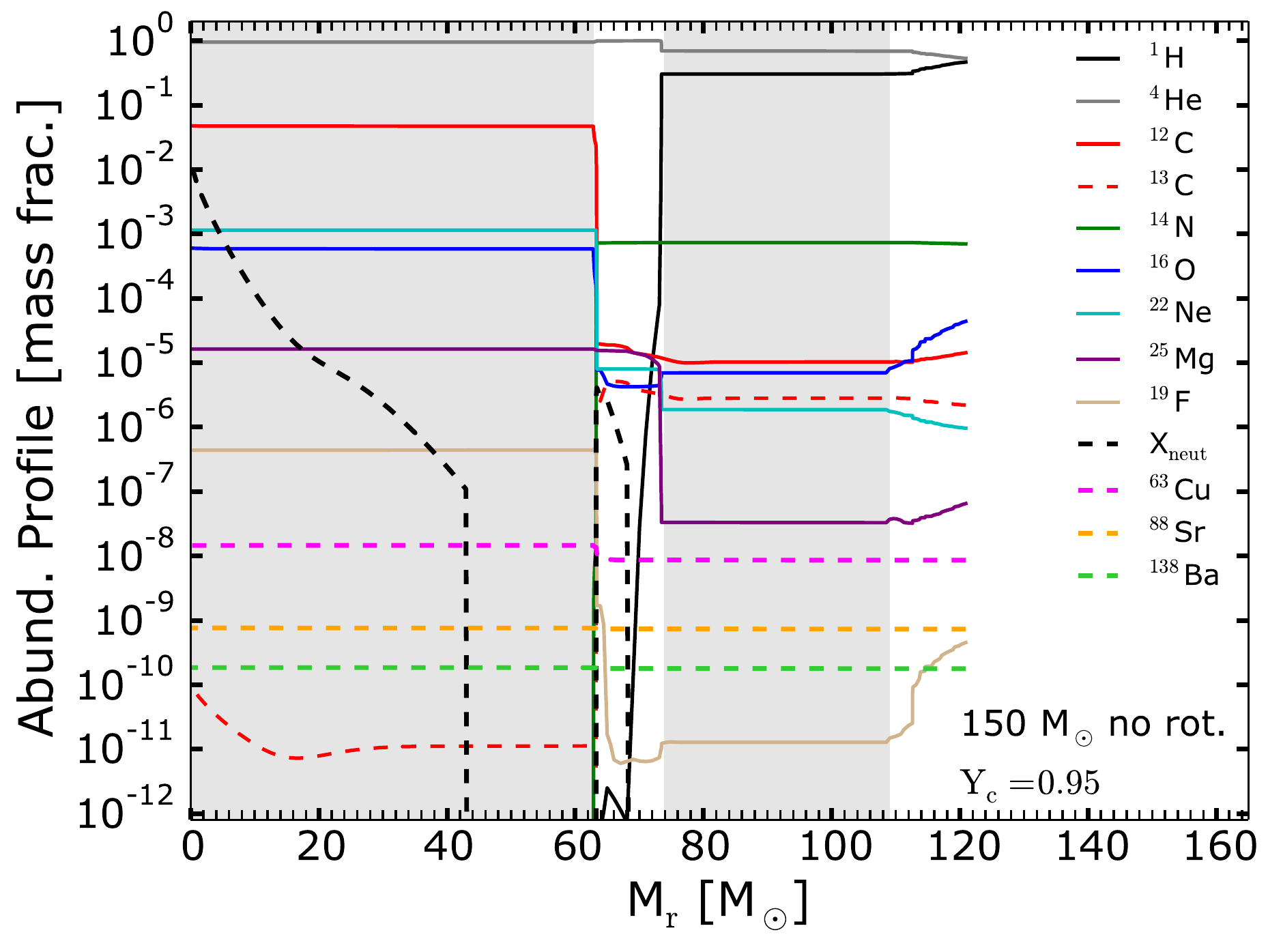}
   \centering
   \end{minipage}
   \begin{minipage}[c]{.49\linewidth}
   \centering
      \includegraphics[scale=0.45]{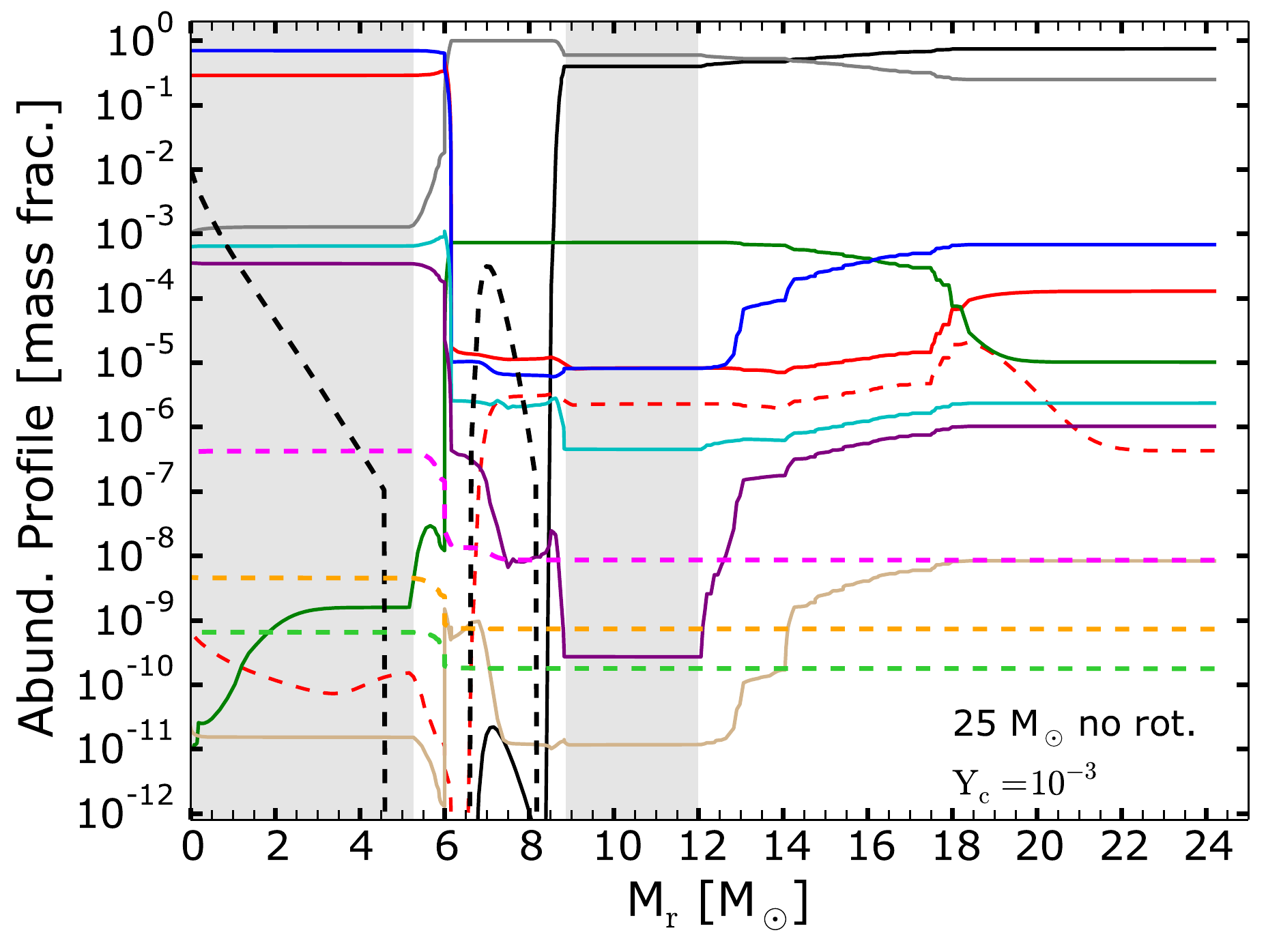}
      \centering
   \end{minipage}
   \begin{minipage}[c]{.49\linewidth}
   \centering
      \includegraphics[scale=0.45]{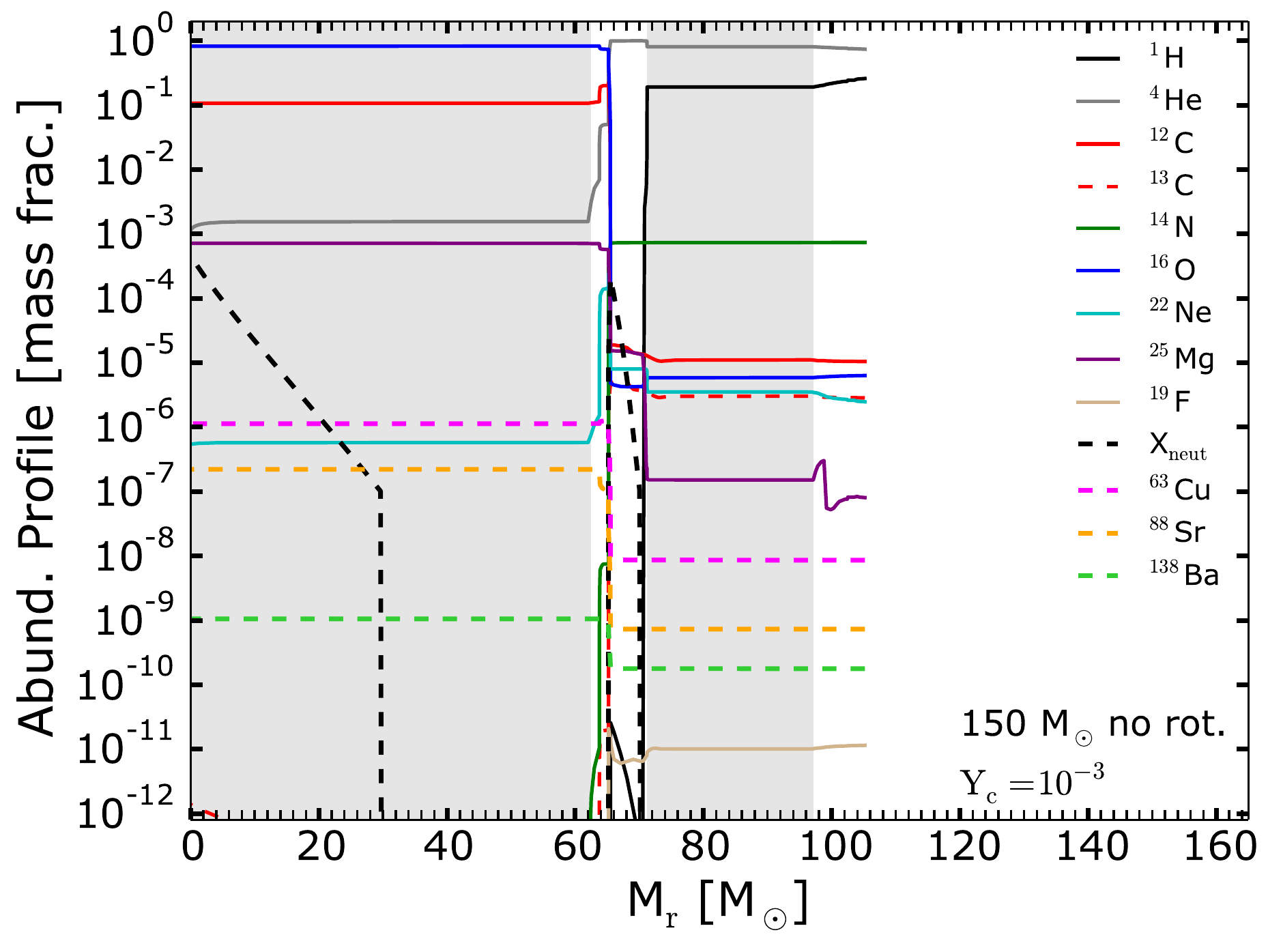}
      \centering
   \end{minipage}
   \caption{Abundance profile of the non-rotating 25 $M_{\odot}$ (left) and 150 $M_{\odot}$ (right) models at the beginning (top panels) and at the end (bottom panels) of central helium burning phase. Grey areas show the convective zones. The neutron profile is scaled up by a factor of $10^{18}$.}
\label{abprof}
    \end{figure*}

\begin{equation}
m_i = \int_{M_{\rm cut}}^{M_{\rm fin}} (X_{\rm i}(M_{\rm r}) - X_{\rm i,0}) \text{d}M_r +  \int_{0}^{\tau} \dot{M}(t) (X_{\rm i,s}(t) - X_{\rm i,0}) \text{d}t,
\label{yie}
\end{equation}
where $M_{\rm{fin}}$ and $\tau$ are the mass at the end of the evolution and the total lifetime of the model, respectively (both given in Table~\ref{table:1}), $X_{\rm i}(M_{\rm r})$ is the mass fraction of isotope $i$ at coordinate $M_{\rm r}$, at the end of the calculation, $X_{\rm i,0}$ is the initial mass fraction, $X_{\rm i,s}(t)$ and $\dot{M}(t)$ are the surface mass fraction and the mass-loss rate at time $t$ respectively. 
As a first step, $M_{\rm cut}$ is estimated using the relation of \cite{maeder92}, that links the mass of the CO-core to the mass of the remnant. Such remnant masses are defined as $M_{\rm rem}$ and are given in the last column of Table~\ref{table:4} for our models\footnote{In \cite{maeder92}, the relation between the mass of the CO-core and the mass of the remnant is applied for $M_{\rm ini} \leq 120$ $M_{\odot}$. For our 150 $M_{\odot}$ model, we have extrapolated the relation linearly.}. 
The impact of different assumptions on the mass cut, and hence on the remnant mass, are discussed in Sect.~\ref{subeffmcut}.

In addition to the yields, we use in this work the productions factors. For an isotope $i$, the production factor is defined as

\begin{equation}
f_i = \frac{M_{\rm ej}}{M_{\rm ini}}\frac{X_{\rm i}}{X_{\rm i, 0}},
\label{fact}
\end{equation}
with $M_{\rm ej}$ the total mass ejected by the star, $M_{\rm ini}$ the initial mass and $X_{\rm i}$ the mass fraction of isotope $i$ in the ejecta. It expresses the ratio of what is given back by the star divided by what was present initially in the whole star.

   \begin{figure*}
   \centering
   \begin{minipage}[c]{.49\linewidth}
      \includegraphics[scale=0.47]{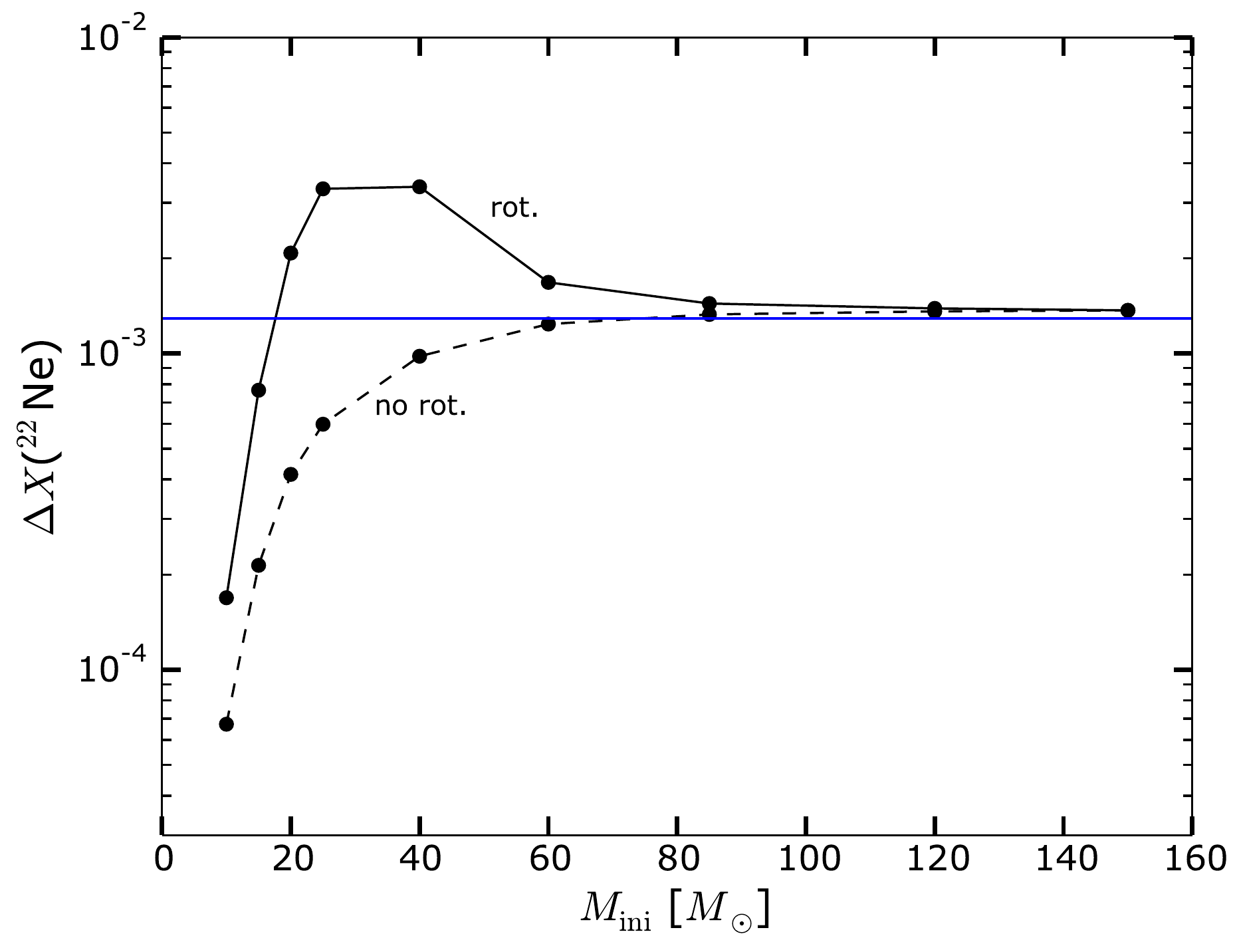}
   \end{minipage}
   \begin{minipage}[c]{.49\linewidth}
      \includegraphics[scale=0.47]{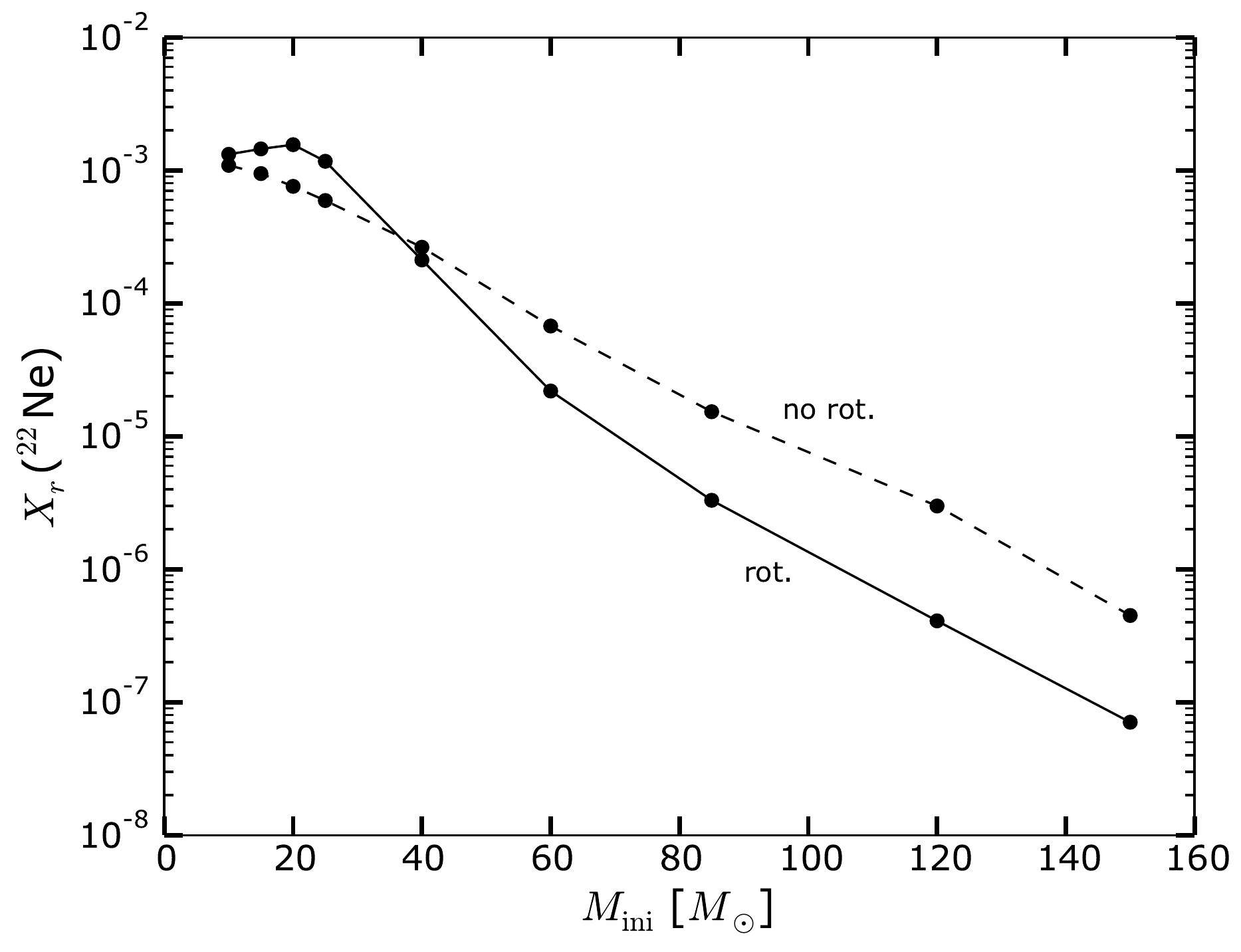}
   \end{minipage}
   \caption{Mass fraction of burnt (left panel) and remaining (right panel) $^{22}$Ne at the end of core helium burning as a function of initial mass. The blue line on the left panel shows the sum of the initial mass fraction of CNO isotopes.}
\label{ne22burnleft}
    \end{figure*}

   \begin{figure*}
   \centering
   \begin{minipage}[c]{.49\linewidth}
      \includegraphics[scale=0.25]{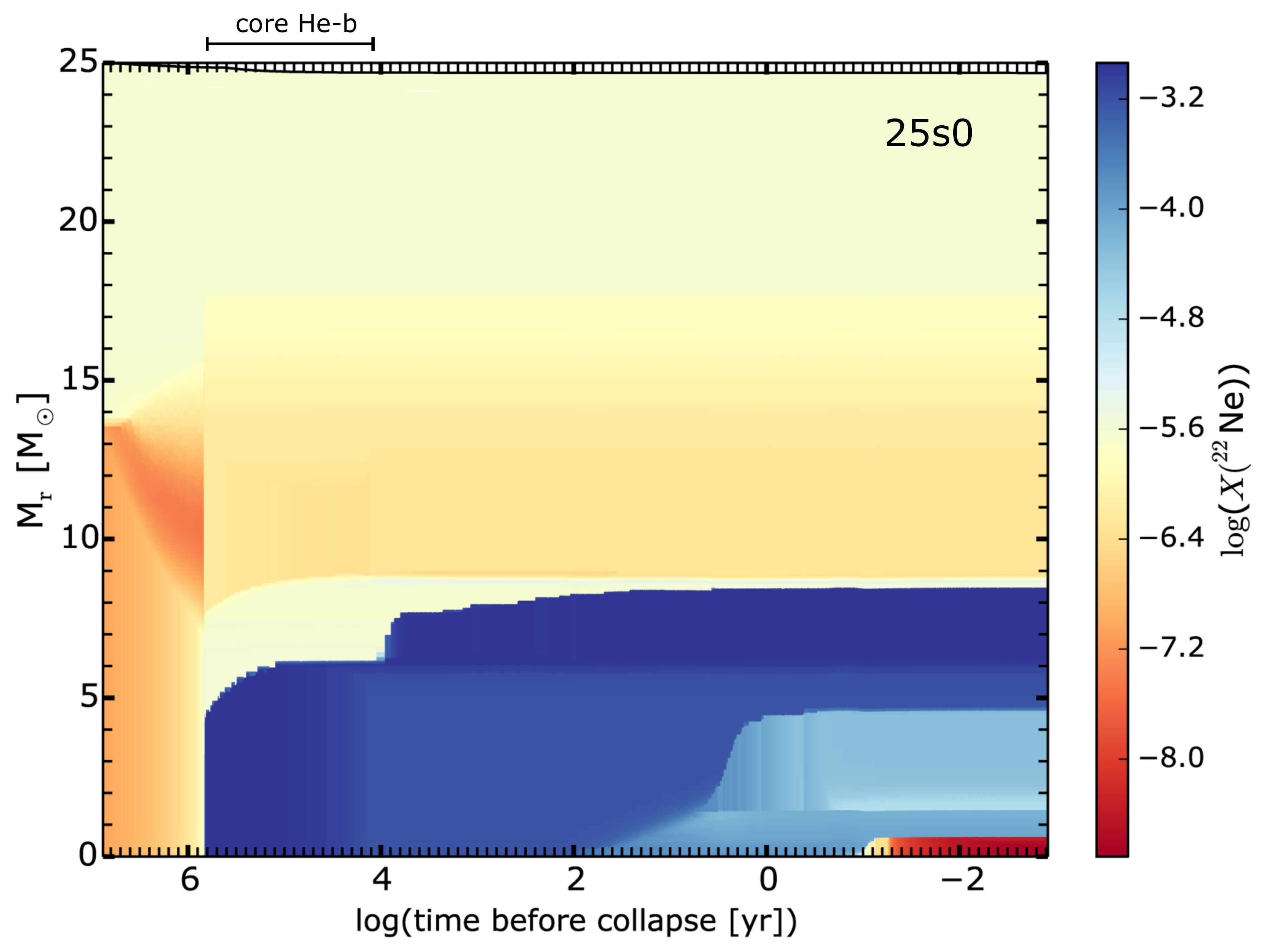}
   \end{minipage}
   \begin{minipage}[c]{.49\linewidth}
      \includegraphics[scale=0.25]{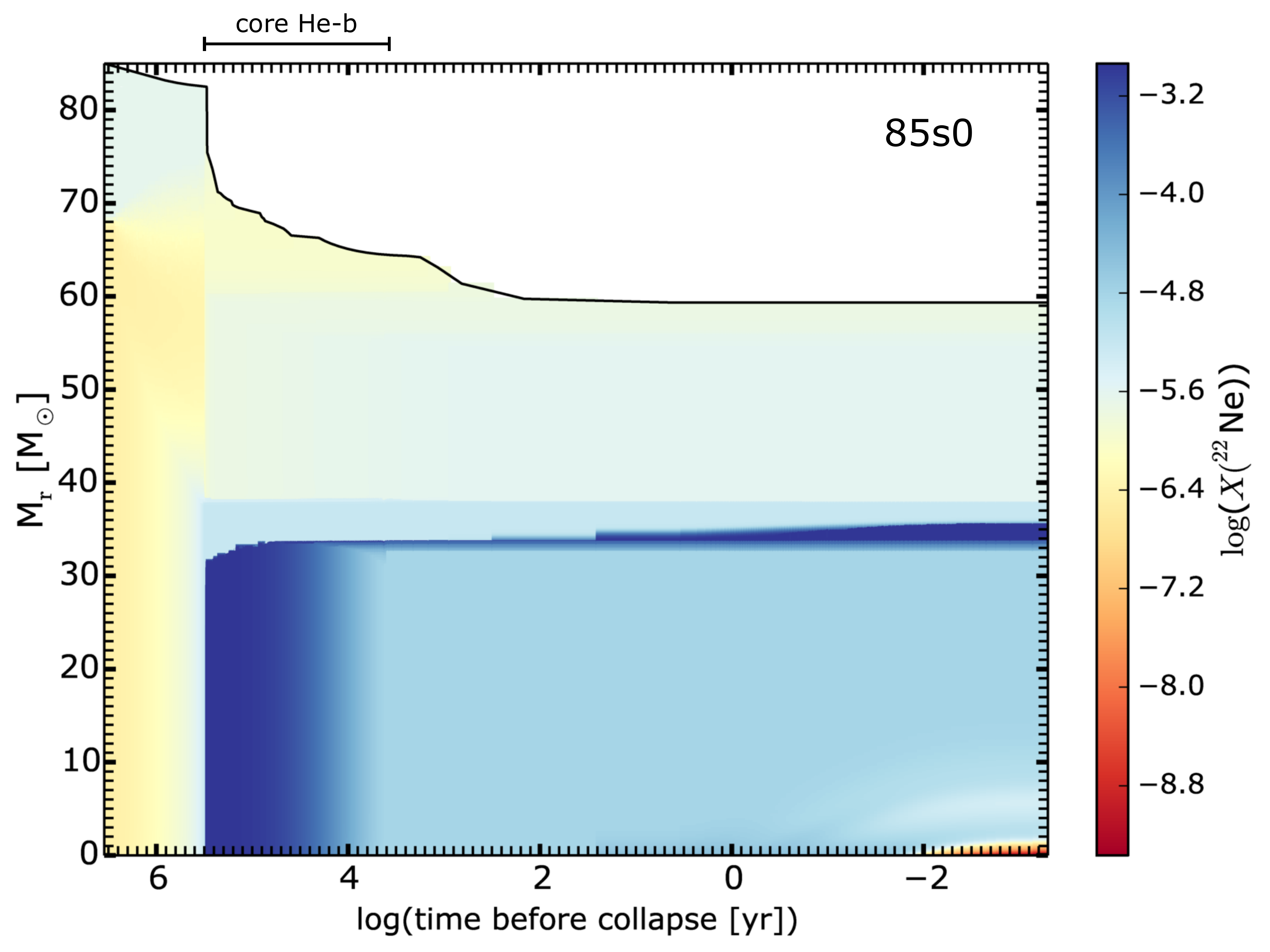}
   \end{minipage}
   \caption{Kippenhahn diagrams of the non-rotating 25 $M_{\odot}$ (left panel) and 85 $M_{\odot}$ (right panel) models. The color map shows the mass fraction of $^{22}$Ne (the initial $^{22}$Ne mass fraction is $\log (X(^{22}$Ne))$_{\rm ini}$ $= -5.6$). The duration of the core helium burning phase is indicated on the top of the panels.}
\label{kip}
    \end{figure*}

\section{Massive stars with rotation and s-process}\label{nuc}

\subsection{Non-rotating models}

The central temperature at the beginning of the helium-burning stage (when the central helium mass fraction $Y_c = 0.95$) is 182, 210 and 220 MK for the 25, 85 and 150 $M_{\odot}$ models, respectively. 
Above 220 MK, the $^{22}$Ne($\alpha,n$)$^{25}$Mg and $^{22}$Ne($\alpha,\gamma$)$^{26}$Ne reactions start to be active and provide the main source of neutrons. Below 220 MK, the neutrons are provided by other ($\alpha,n$) reactions from elements between C and Ne (especially $^{13}$C($\alpha,n$)$^{16}$O). 
The top panels of Fig.~\ref{abprof} show the neutron profile (black dashed lines) in the 25 and 150 $M_{\odot}$ models at the beginning of the core helium burning phase. 
The central neutron peak is bigger for the 150 $M_{\odot}$ because of the higher central temperature that activates more efficiently the ($\alpha,n$) reactions between N and Ne.
At this early stage of core He-burning, the s-process is not activated significantly (see the flat $^{88}$Sr and $^{138}$Ba profiles in the top panels of Fig.~\ref{abprof}) and leads only to slight overabundances of light s-elements like $^{63}$Cu (dashed magenta line).

At the end of core helium burning (bottom panels), the temperature $T>220$ MK in the core so that the main neutron source in the He-burning core is $^{22}$Ne for both models.
Also, in both models and during all the core He-burning phase, the second neutron peak (at higher mass coordinates) is mainly due to $^{13}$C($\alpha,n$). The s-process is not efficient in this region (see the $^{63}$Cu, $^{88}$Sr and $^{138}$Ba profiles) because of the high $^{14}$N abundance, acting as a strong neutron poison.

In more massive models, the temperature required for the efficient activation of the $^{22}$Ne($\alpha,n$) reaction (220 MK) is reached earlier during the core helium burning phase: while the 150 $M_{\odot}$ model reaches a central temperature $T_c = 220$ MK at the very start of core He-burning, the 25 $M_{\odot}$ model reaches this temperature only close to the end of He-burning, when $Y_c \sim 0.2$.
The duration of the stage where the central temperature $T_c > 220$ MK is 0.16, 0.22 and 0.25 Myr for the 25, 85 and 150 $M_{\odot}$ models, respectively. 
The amount of burnt $^{22}$Ne during core He-burning therefore increases with initial mass (dashed line in the left panel of Fig.~\ref{ne22burnleft}).
As the initial mass increases, it converges toward a plateau whose value is almost equal to the sum of the initial CNO mass fraction X(CNO)$_{\rm ini}$ (horizontal blue line in the left panel of Fig.~\ref{ne22burnleft}). 
Indeed, during the Main Sequence, the CNO cycle mainly transforms $^{12}$C and $^{16}$O into $^{14}$N. Consequently, at the end of the Main Sequence, X(CNO)$_{\rm ini} \simeq$ X($^{14}$N) in the core. When the core helium burning phase starts, $^{14}$N is transformed into $^{22}$Ne by successive $\alpha$ captures. 
Hence, at core He depletion, the maximum amount of burnt $^{22}$Ne is about X(CNO)$_{\rm ini}$.

Since more $^{22}$Ne is burnt in more massive stars, less is left at core He depletion (e.g. Fig.~\ref{kip} and the dashed line in the right panel of Fig.~\ref{ne22burnleft}). For stars with $M_{\rm ini} > 40$ $M_{\odot}$, almost all the available $^{22}$Ne burns during the core helium burning phase so that the contribution of the C-shell burning in producing s-element is in general negligible. 
Additional contributions from $^{13}$C($\alpha,n$)\footnote{starting from $^{12}$C($p,\gamma$), with
the protons coming from $^{12}$C($^{12}$C$,p$)$^{23}$Na.} or $^{12}$C($^{12}$C$,n$)$^{23}$Mg are in principle possible during carbon burning \citep[see][for more details]{bennett12,pignatari13} but these contributions generally remain much smaller than the $^{22}$Ne contribution during He-burning for very massive stars.

   \begin{figure*}[t]
   \centering
      \includegraphics[scale=0.62]{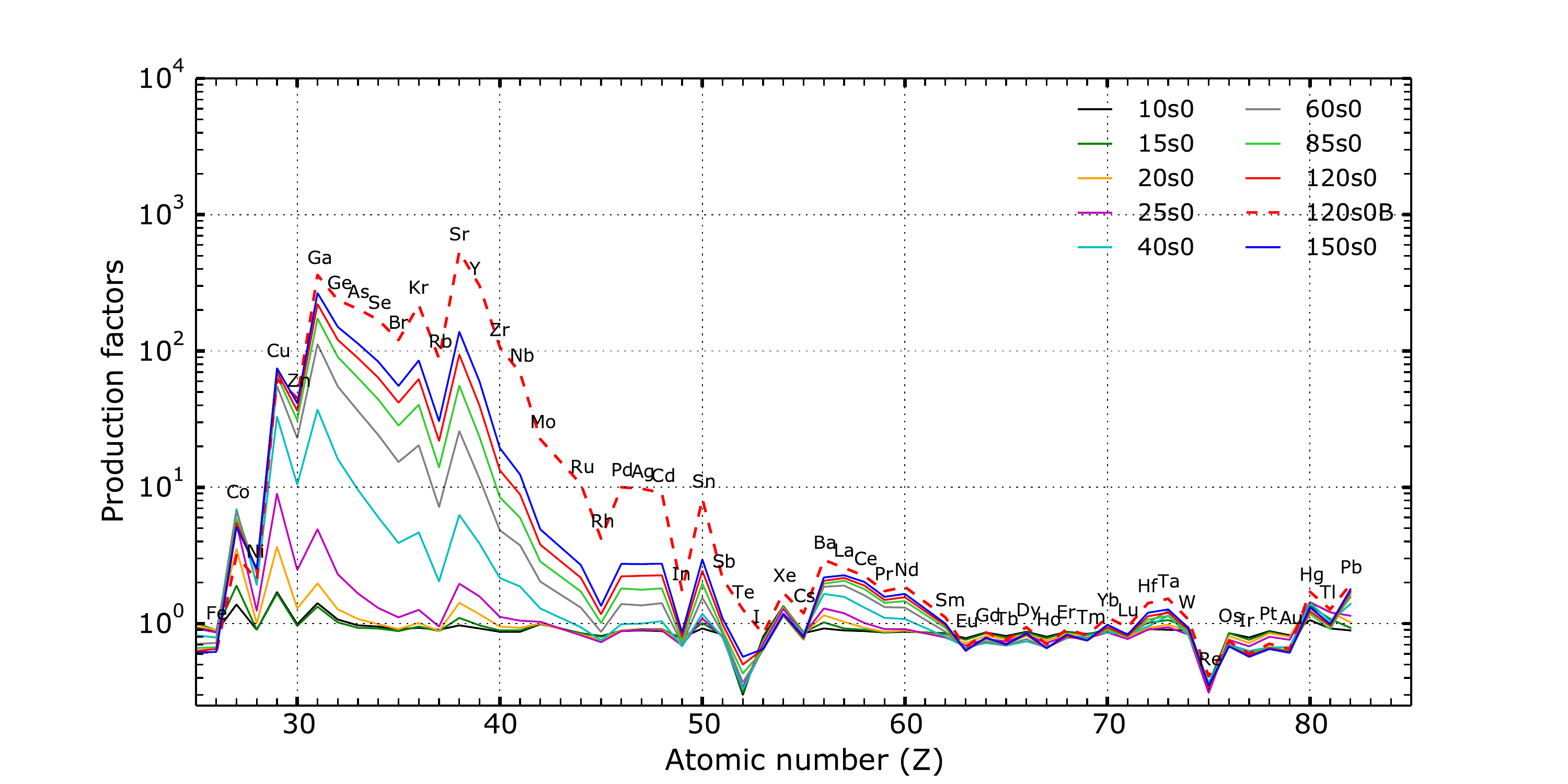}
   \caption{Production factors (Eq.~\ref{fact}) of non-rotating models. The mass cut is set according to the relation of \cite{maeder92}.}
\label{zm3norot}
    \end{figure*}

   \begin{figure*}
   \centering
      \includegraphics[scale=0.62]{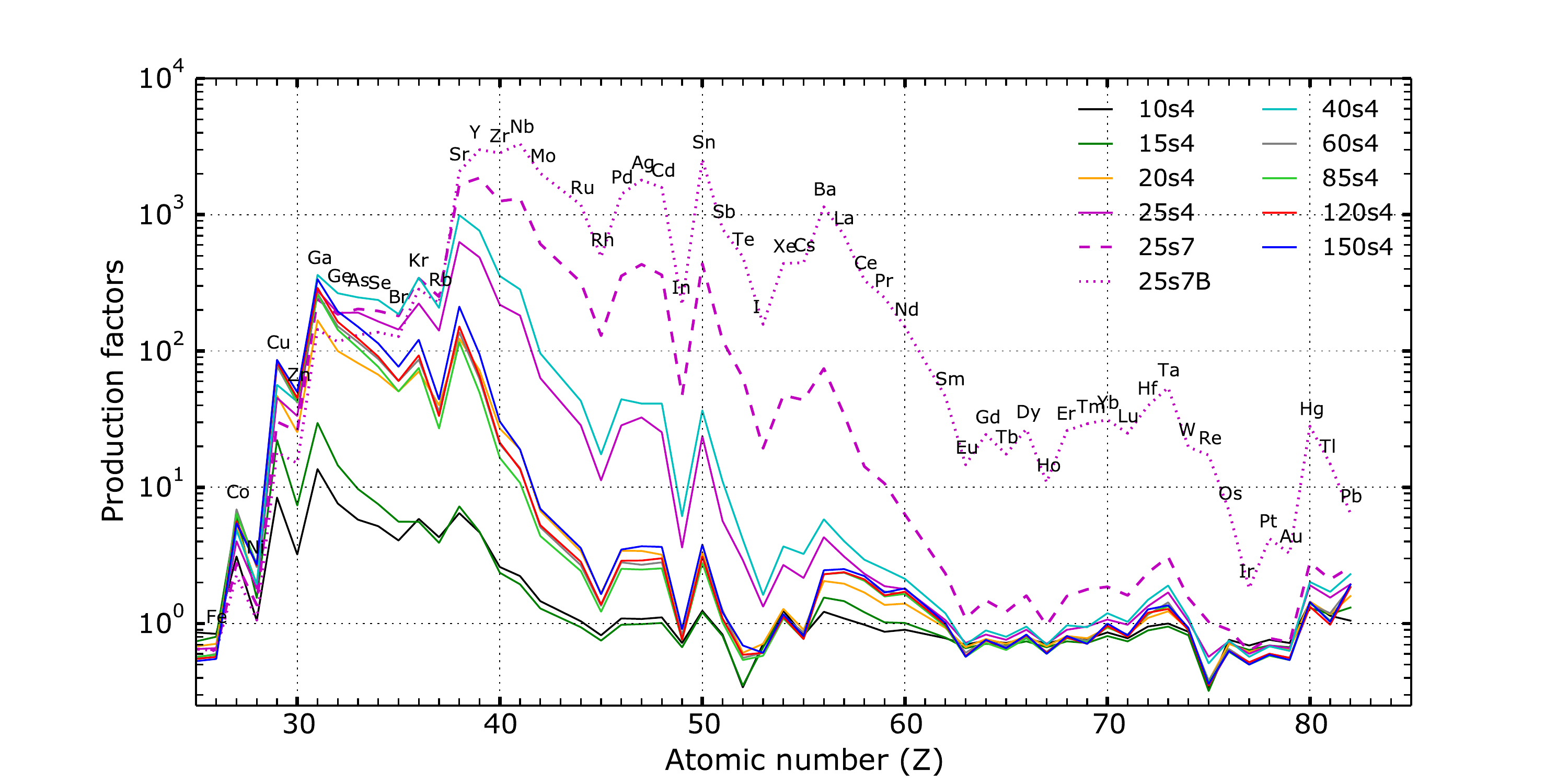}
   \caption{Same as Fig.~\ref{zm3norot} but for rotating models.}
\label{zm3rot}
    \end{figure*}

A higher temperature also favors the production of s-elements because the ratio of the rate of $^{17}$O($\alpha,n$)$^{20}$Ne over the rate of $^{17}$O($\alpha,\gamma$)$^{21}$Ne increases with increasing temperature. It means that for higher temperatures, the poisoning effect of $^{16}$O is reduced since neutrons are more efficiently recycled\footnote{$^{16}$O is an abundant poison that absorbs neutrons in the He-core and limits the production of s-elements. With the chain $^{16}$O($n,\gamma$)$^{17}$O($\alpha,\gamma$)$^{21}$Ne, the neutron captured by $^{16}$O is definitely lost. With the chain $^{16}$O($n,\gamma$)$^{17}$O($\alpha,n$)$^{20}$Ne, the neutron is captured by $^{16}$O and then recycled.} by $^{17}$O. The mean central temperature of the 25 and 150 $M_{\odot}$ models during He-burning are 207 and 233 MK respectively. At 233 MK, the ratio ($\alpha,n$) / ($\alpha,\gamma$) is roughly twice that at 207 MK.

Finally, s-elements are also overproduced in more massive stars because these stars have larger He-burning cores. The mass of the He-burning core corresponds roughly to the mass of the CO-core at the end of the evolution, which increases with initial mass (column 6 in Table~\ref{table:4}) and also represents a larger fraction of the final stellar mass. 

The effects discussed above lead to an increasing production of s-elements with initial stellar mass (Fig.~\ref{zm3norot}), as was already found before \citep{langer89,prantzos90,kappeler99,the07}. In our models, the production factors of the s-elements for the 150 $M_{\odot}$ model exceed the factors of the 25 $M_{\odot}$ model by about 2 orders of magnitude at maximum (e.g. for Sr, see Fig.~\ref{zm3norot}). Whatever the mass, the production factors of elements with $Z \gtrsim 50$ stay very small. Considering a lower $^{17}$O($\alpha,\gamma$)$^{21}$Ne rate mostly affects the range $30 \lesssim Z \lesssim 50$ (see red dashed line on Fig.~\ref{zm3norot}). 
Even if the production factors of heavy s-elements like Pb do not vary much (a factor of $\sim 2$) in the considered mass range, the Pb yield in $M_{\odot}$ (Eq.~\ref{yie}) is about 3 dex higher in the 150 $M_{\odot}$ compared to the 10 $M_{\odot}$, because much more mass (hence Pb) is ejected in the case of the 150 $M_{\odot}$.

   \begin{figure*}
   \centering
      \includegraphics[scale=0.79]{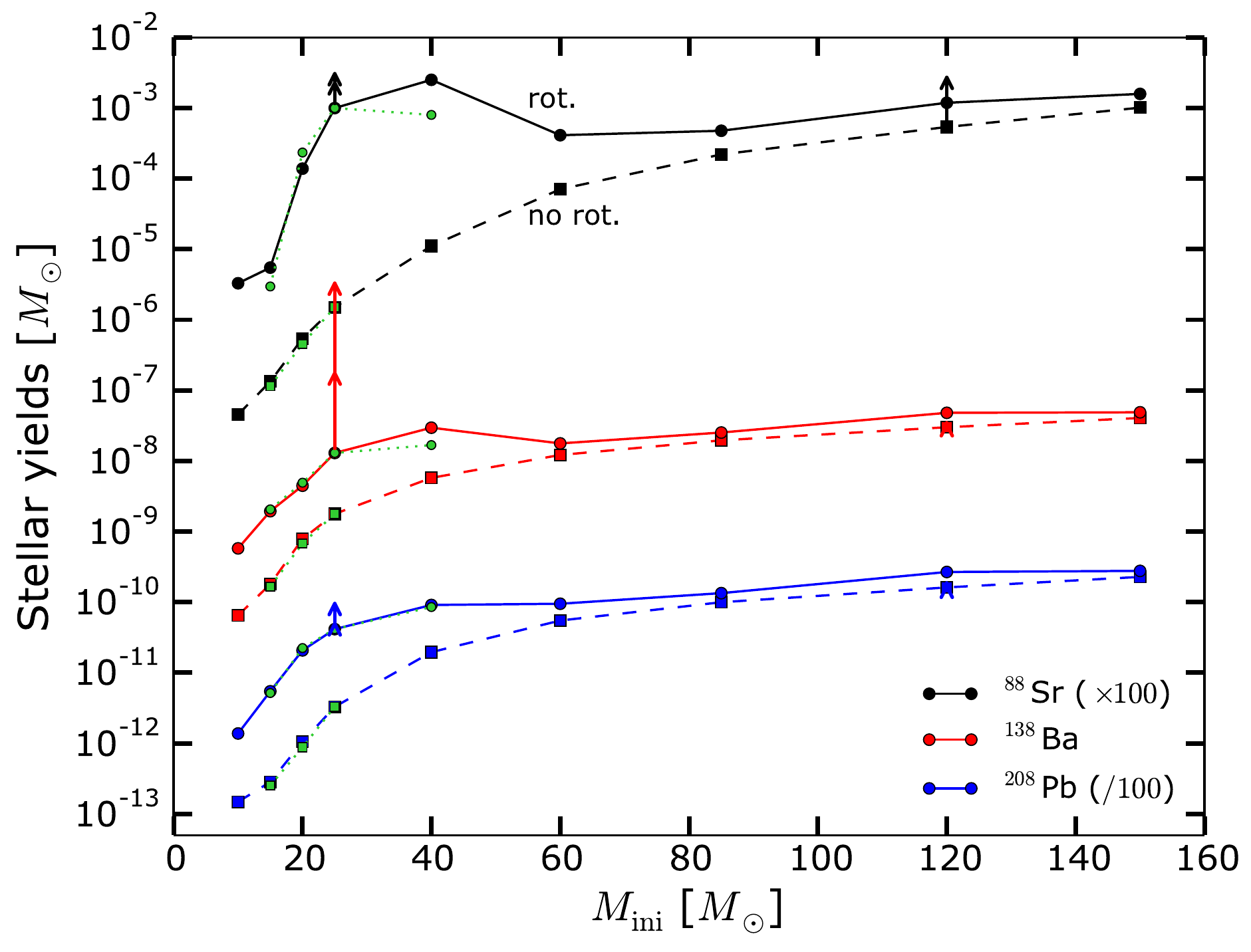}

\caption{Stellar yields in $M_{\odot}$ (Eq.~\ref{yie}) as a function of the initial mass $M_{\rm ini}$ for the non-rotating (dashed lines) and rotating (solid lines) models. The mass cut is set using the relation of \cite{maeder92}. $^{88}$Sr, $^{138}$Ba and $^{208}$Pb isotopes are shown. Green patterns between 15 and 40 $M_{\odot}$ are the yields of the F16 models. The small and big red arrows at $M_{\rm ini}=25$ $M_{\odot}$ indicate the yields of $^{138}$Ba for the fast rotating 25 $M_{\odot}$ model and the fast rotating 25 $M_{\odot}$  models with a lower $^{17}$O($\alpha,\gamma$) rate respectively. The same arrows are plotted for $^{88}$Sr and $^{208}$Pb. Arrows at $M_{\rm ini}=120$ $M_{\odot}$ represent the 120 $M_{\odot}$ model with a lower $^{17}$O($\alpha,\gamma$) rate. Note that some arrows are not visible because there are too small.}
            
  \label{yields}
    \end{figure*}

\subsection{Rotating models}\label{rotmodels}

The boost on s-elements production due to rotation is the highest between 20 and 60 $M_{\odot}$. Fig.~\ref{zm3rot} shows indeed that the production factors first increase from 10 to 40 $M_{\odot}$ and decrease for $M>40$ $M_{\odot}$. As shown in Fig.~\ref{zm3rot}, the 60, 85, 120, and 150 $M_{\odot}$ models with rotation have similar patterns. 
Fig.~\ref{yields} shows that our models agree well with the 15, 20, 25 and 40 $M_{\odot}$ of F16 (green pattern). It means that overall, the new rates used in the present work (c.f. Table~\ref{table:2}) do not affect much the yields compared to the yields published in F16.
Fig.~\ref{yields} also shows that rotation affects significantly the yields only if $M_{\rm ini} < 60$ $M_{\odot}$.
For $10 < M_{\rm ini} < 40$ $M_{\odot}$, the $^{88}$Sr, $^{138}$Ba and $^{208}$Pb yields are boosted by $\sim 2-3$, $\sim 1$ and $\sim 1$ dex, respectively (see Fig.~\ref{yields}). 
There are two main reasons for that:

   \begin{figure*}
   \begin{minipage}[c]{.49\linewidth}
   \centering
      \includegraphics[scale=0.43]{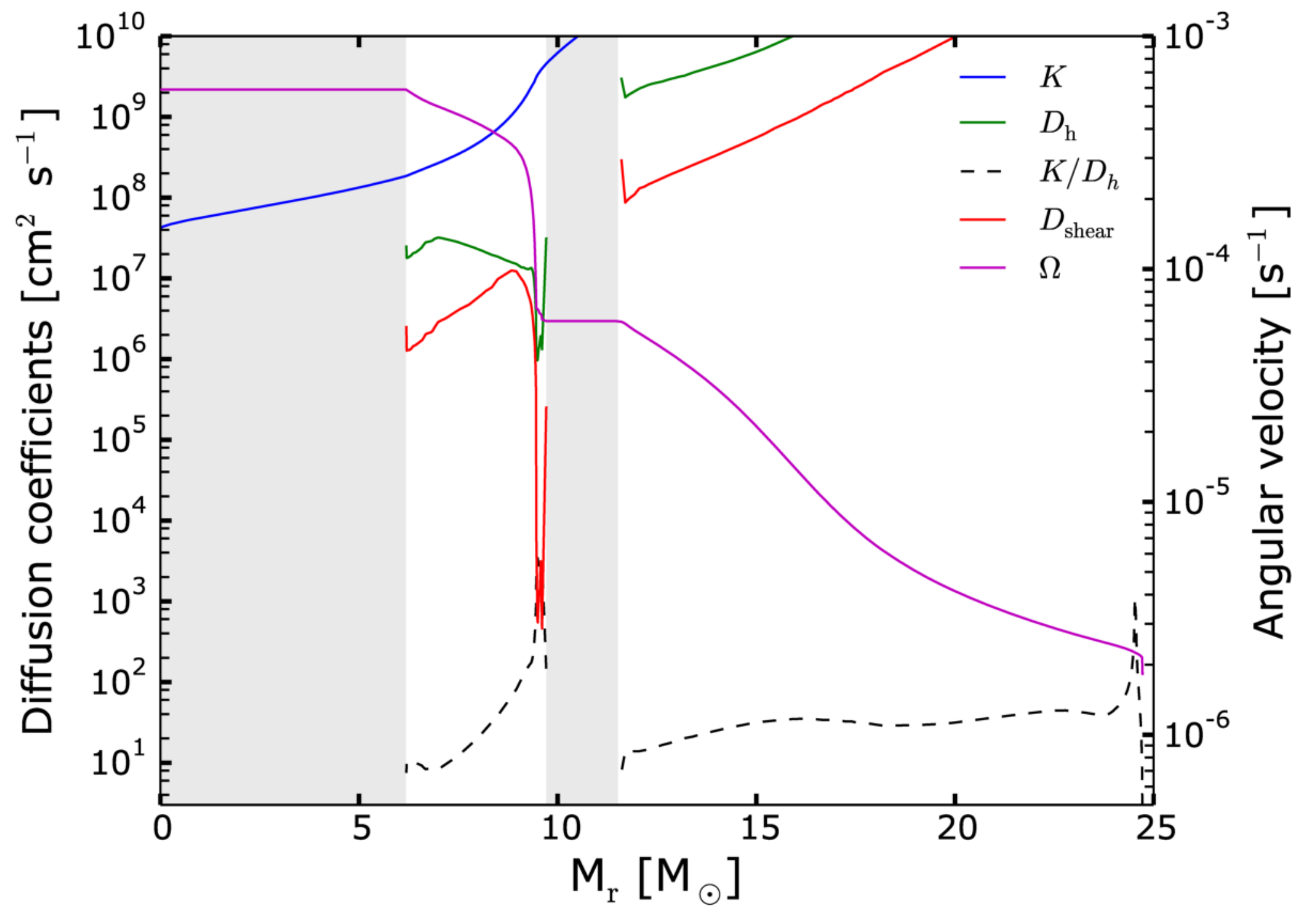}
      \centering
   \end{minipage}
   \begin{minipage}[c]{.49\linewidth}
   \centering
      \includegraphics[scale=0.43]{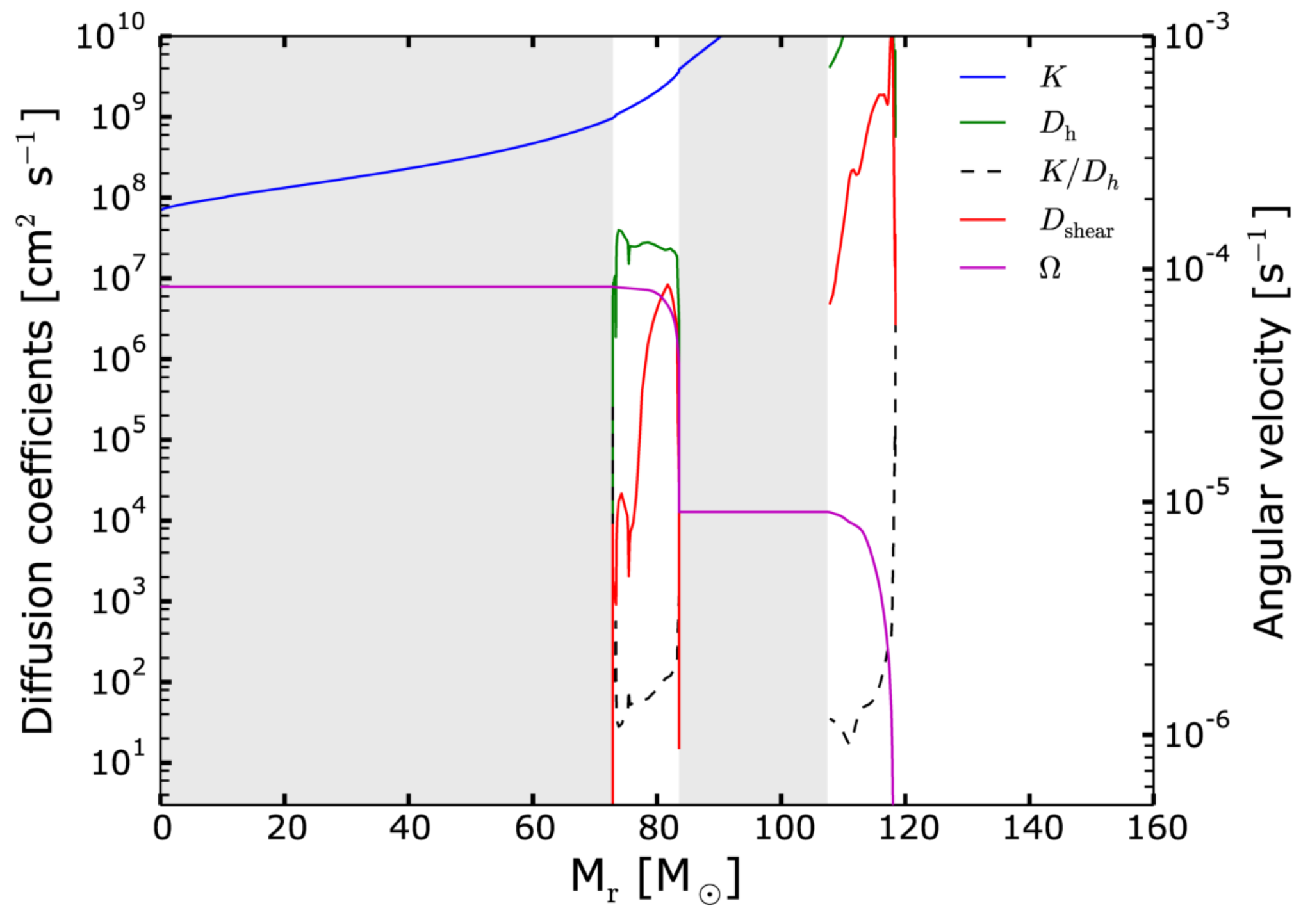}
      \centering
   \end{minipage}
   \caption{$\Omega$ profile, $D_{\rm shear}$ coefficient and other diffusion coefficients in Eq.~\ref{dshtz97} for the rotating 25 (left) and 150 $M_{\odot}$ models (right) during the core He-burning phase ($Y_c = 0.66$). Grey areas represent the convective zones and the dashed line show the $K/D_{\rm h}$ ratio.}
\label{coef}
    \end{figure*}

   \begin{figure*}[t]
   \centering
   \begin{minipage}[c]{.49\linewidth}
   \centering
      \includegraphics[scale=0.45]{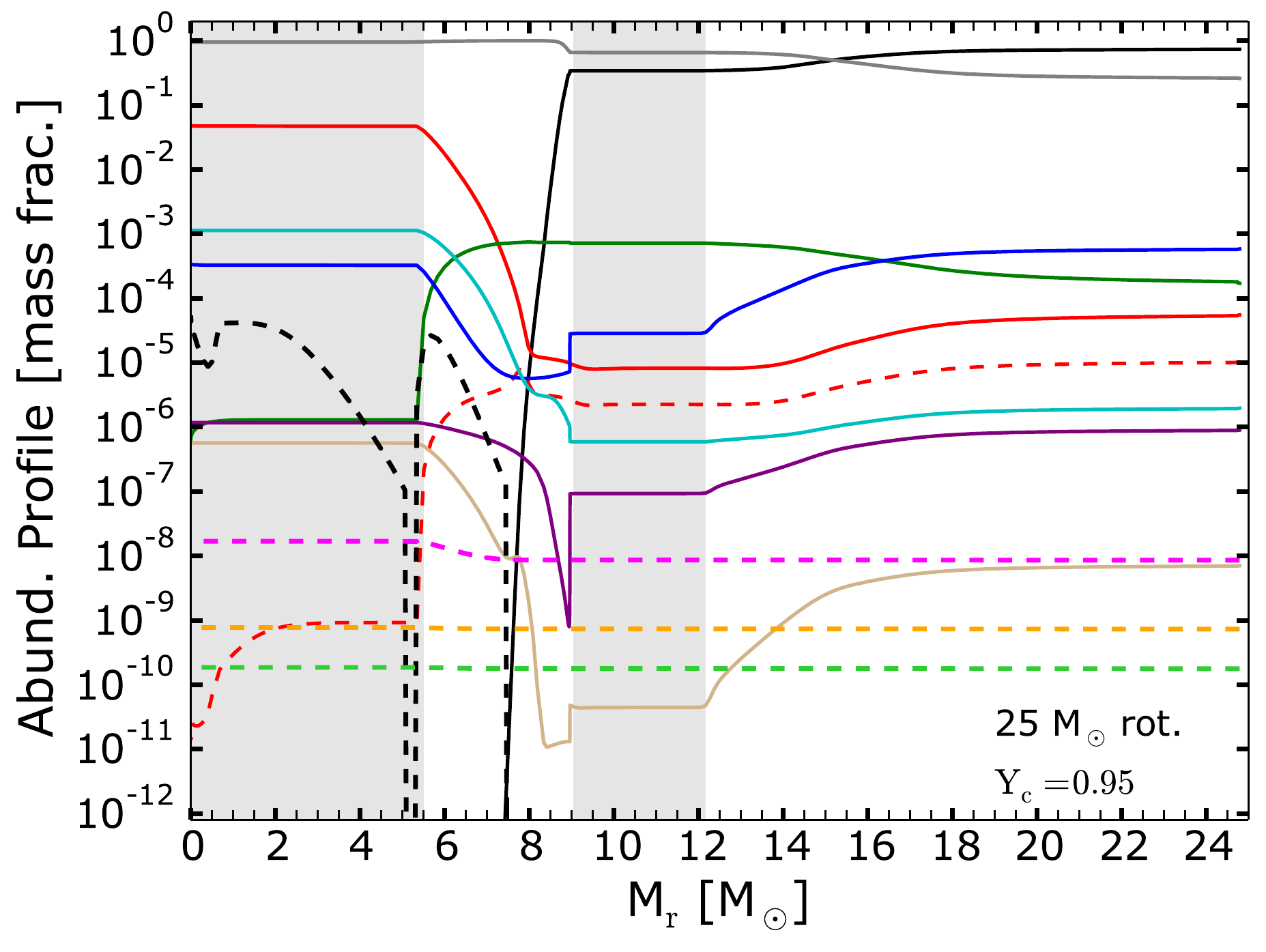}
      \centering
   \end{minipage}
   \begin{minipage}[c]{.49\linewidth}
   \centering
      \includegraphics[scale=0.45]{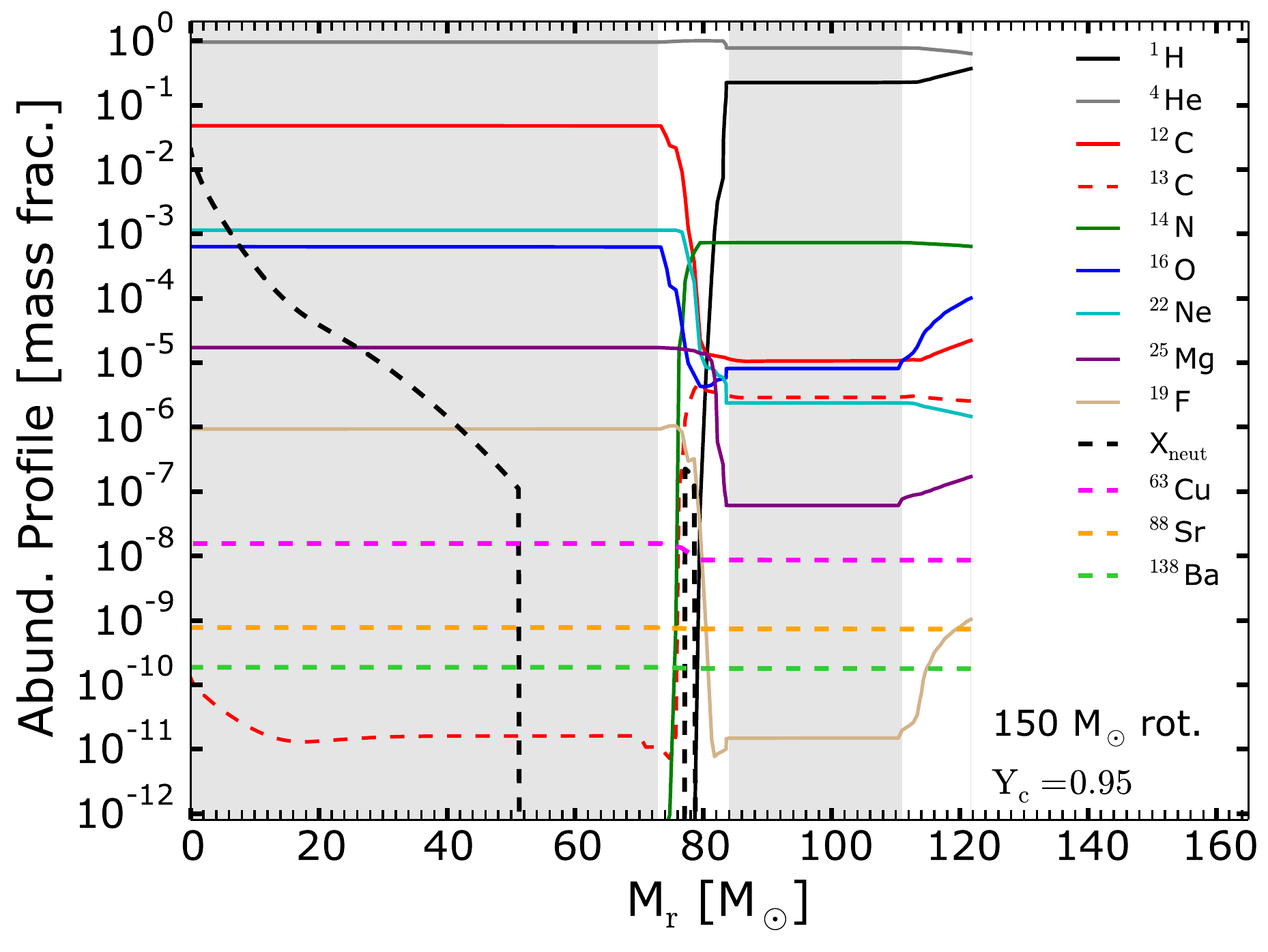}
      \centering
   \end{minipage}
   \begin{minipage}[c]{.49\linewidth}
   \centering
      \includegraphics[scale=0.45]{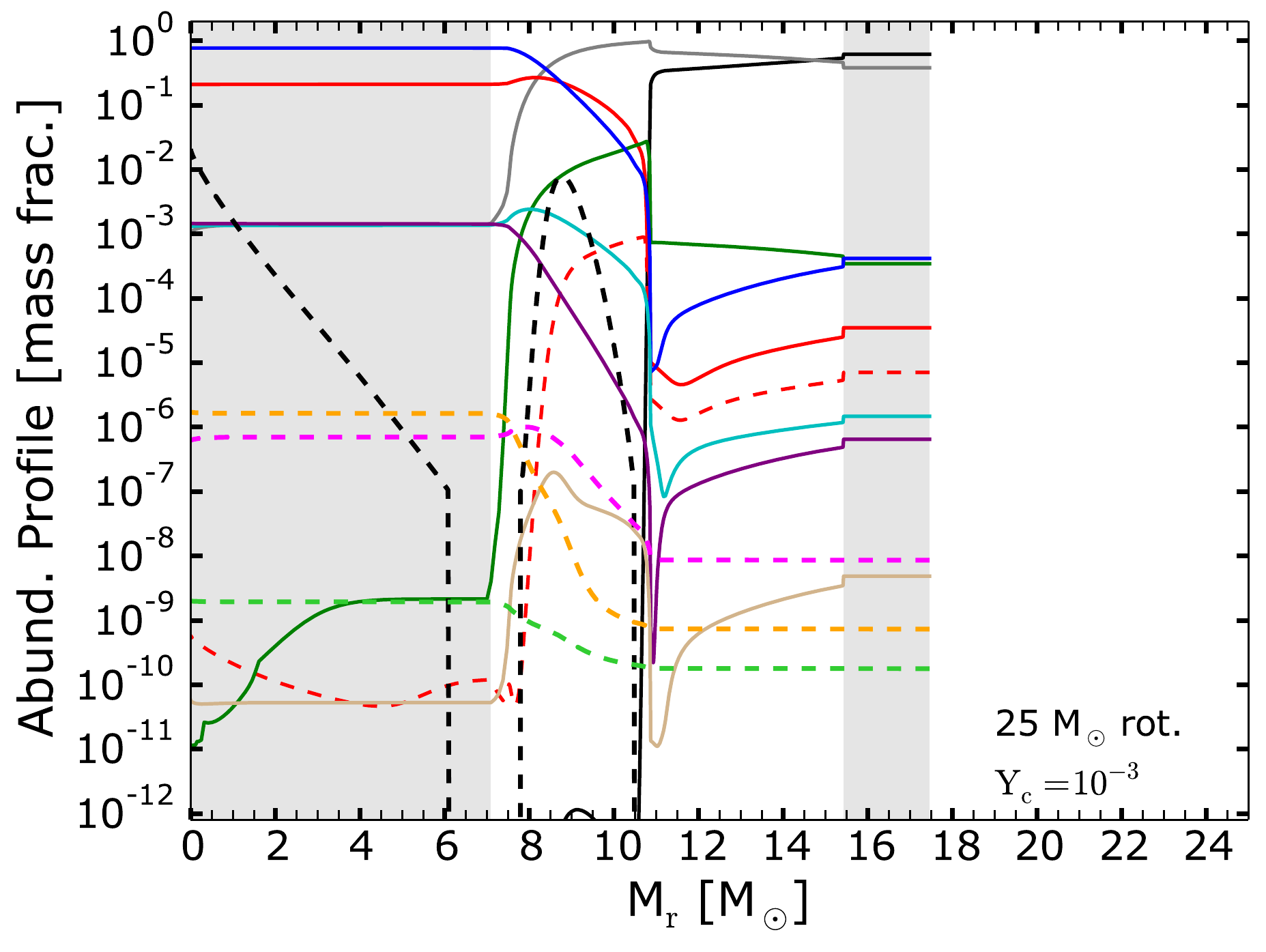}
      \centering
   \end{minipage}
   \begin{minipage}[c]{.49\linewidth}
   \centering
      \includegraphics[scale=0.45]{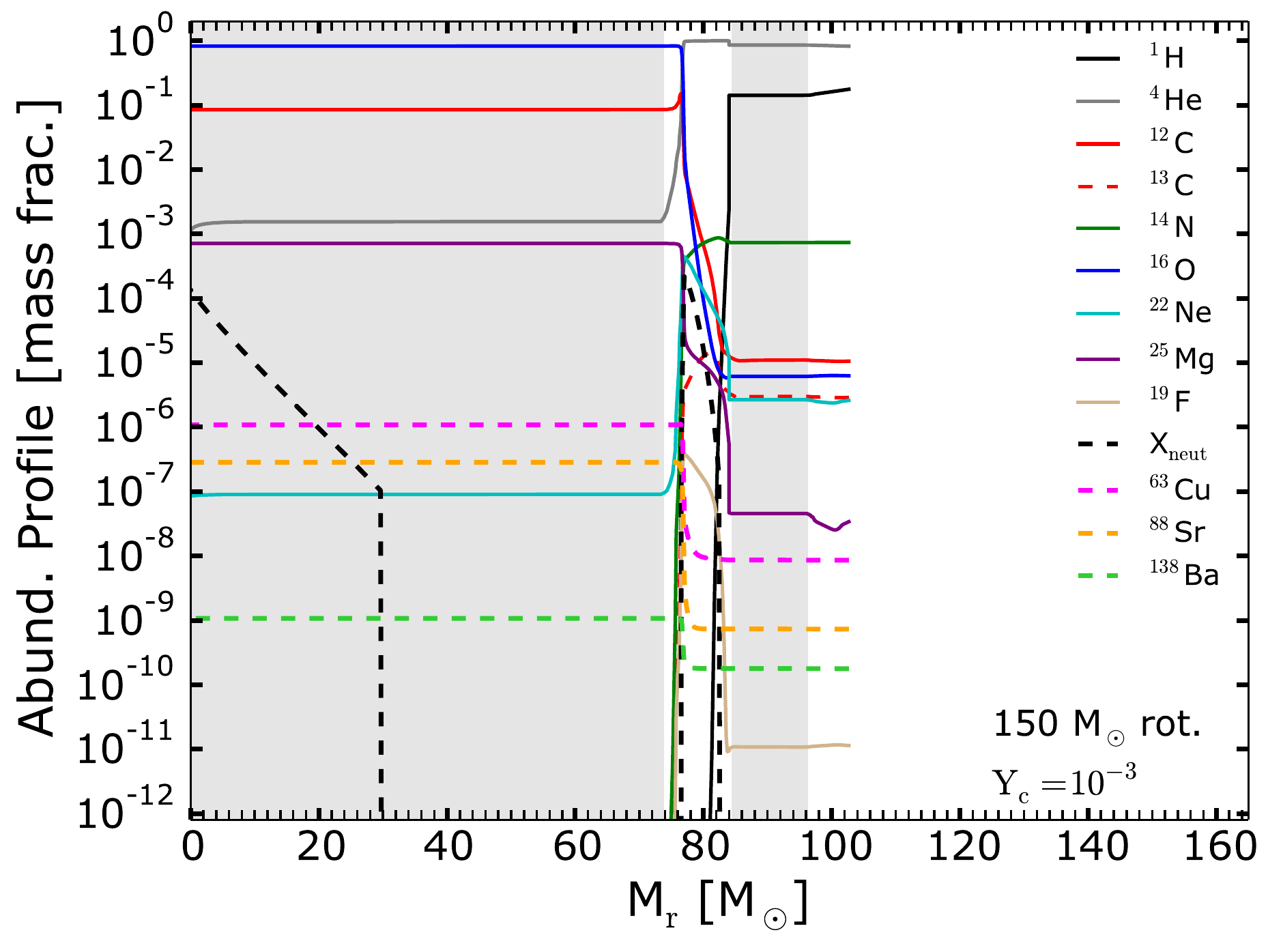}
      \centering
   \end{minipage}
   \caption{Abundance profile of the rotating 25 $M_{\odot}$ (left) and 150 $M_{\odot}$ (right) models at the beginning (top panels) and at the end (bottom panels) of central helium burning phase. Grey areas show the convective zones. The neutron profile is scaled up by a factor of $10^{18}$.}
\label{abprof2}
    \end{figure*}

\begin{itemize}

\item [$\bullet$] First, close to the convective helium burning core, the $D_{\rm shear}$ coefficient (Eq.~\ref{dshtz97}) is generally smaller in more massive models (Fig.~\ref{coef}, red line between the 2 convective zones) and hence transports less efficiently the He-burning products to the H-shell. It ultimately leads to a smaller amount of extra $^{22}$Ne, hence less neutrons. The smaller $D_{\rm shear}$ is explained by the fact that (1) more massive stars have higher $K/D_{\rm h}$ ratios just above the convective core (dashed line in Fig.~\ref{coef}, this point was already discussed in F16, Sect.~3.1) and (2) $\Omega$ and its gradient are smaller in this same region (magenta line). Also, for the 25 $M_{\odot}$ (left panel of Fig.~\ref{coef}) the $D_{\rm shear}$ drops just below the convective H-shell. This is because the convective H-shell migrates upward and leaves behind an almost flat $\Omega$ profile (at $M_{\rm r} \sim 10$ $M_{\odot}$) which strongly reduces the $D_{\rm shear}$ (see also F16, Sect.~3.1). However, the bottom of the H-envelope extends down to about 9 $M_{\odot}$ so that the He-products reach the H-rich region anyway and extra $^{14}$N can be synthesized.

\item [$\bullet$] The second reason is that the most massive stars have a more active H-burning shell. The shell remains convective during the whole He-burning stage and contributes well to the total stellar luminosity. This limits the growth of the He-core of the most massive stars. The growth of the convective He-burning core contributes to form extra $^{22}$Ne by engulfing $^{14}$N. Since the He-core of the most massive stars does not grow as much as the core of less massive stars, less primary $^{14}$N is engulfed in the He-core, leading to a smaller production of s-elements. A word of caution is here required: 
the previous statement may be affected by the current uncertainties in convective boundaries and the mixing across it. For example, using Ledoux criterion instead of Schwarzschild criterion may limit the extent and growth of both the convective H-burning shell and He-burning core. These uncertainties can be tackled with multi-dimensional hydrodynamic simulations and asteroseismology \citep[e.g.][]{arnett15,arnett17, cristini17}.

\end{itemize}

For these reasons, more extra $^{22}$Ne is available and burnt in $M_{\rm ini}<60$ $M_{\odot}$ models, as shown by the bump between 20 and 60 $M_{\odot}$ in Fig.~\ref{ne22burnleft} (solid line, both panels). 
The bottom panels of Fig.~\ref{abprof2} show the abundance profiles at the end of the core He-burning phase for the rotating 25 $M_{\odot}$ (left) and 150 $M_{\odot}$ (right) models. We see indeed that less primary $^{14}$N is synthesized in the 150 $M_{\odot}$ model (compare the $^{14}$N bumps at $M_r$ $\sim 9$ and $80$ $M_{\odot}$ for the 25 and 150 $M_{\odot}$ models respectively).

\begin{figure}[h!]
   \centering
      \includegraphics[scale=0.44]{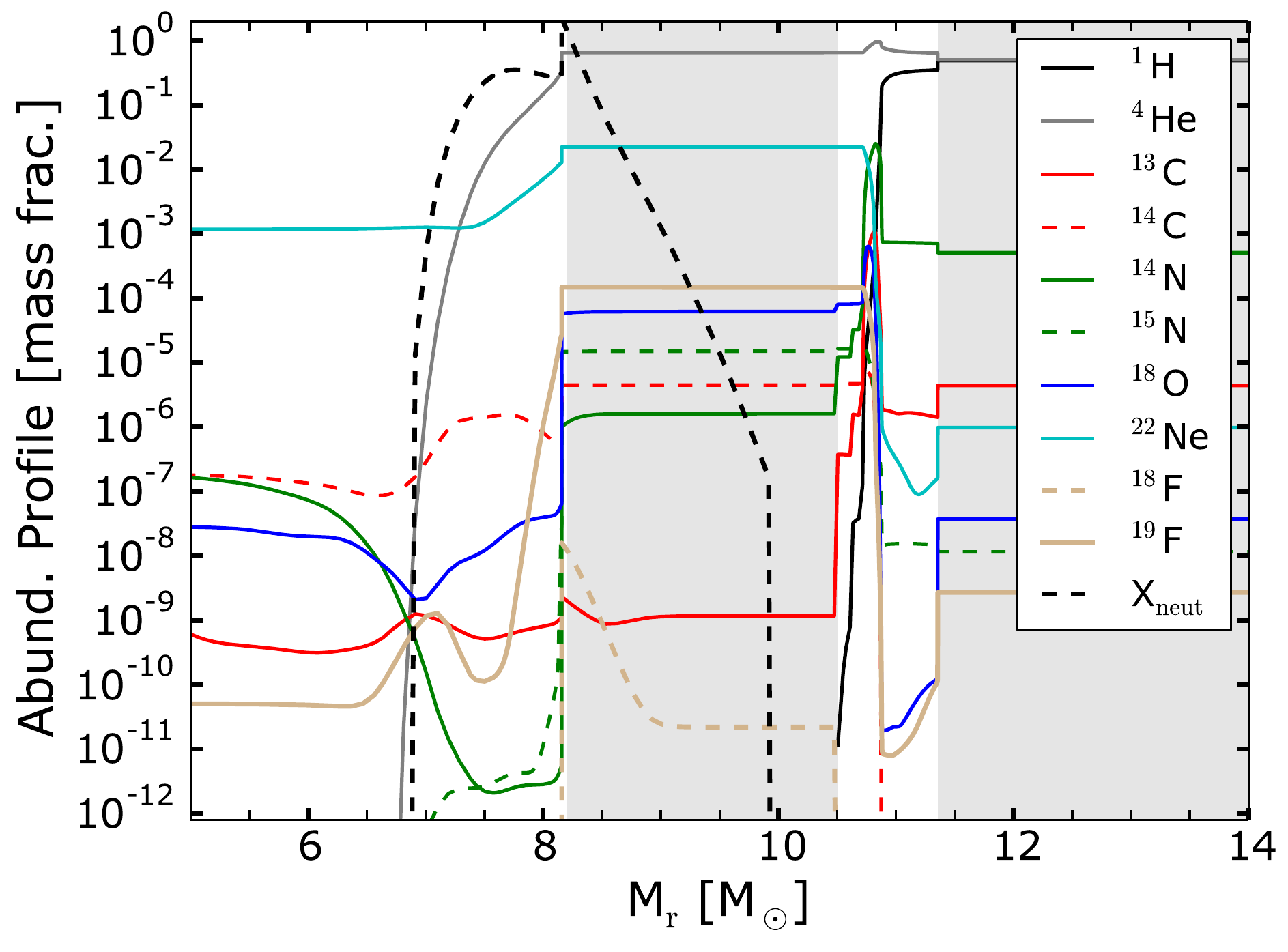}
      \centering
   \caption{Abundance profile of the rotating 25 $M_{\odot}$ during the shell He-burning phase. Grey areas show the convective zones (the convective He-burning shell is in between $\sim 8$ and $\sim 10.5$ $M_{\odot}$). The neutron profile is scaled up by a factor of $10^{18}$.}
\label{f19}
    \end{figure}

$^{19}$F is also an important product of rotation, which is synthesized after the core He-burning phase, in the He-burning shell. Fig.~\ref{f19} shows the convective He-burning shell (between $\sim 8$ and $\sim 10.5$ $M_{\odot}$) of the rotating 25 $M_{\odot}$ model. The abundance of $^{19}$F is about $10^{-4}$ in the He-shell (t is only about $10^{-7}$ in the non-rotating 25 $M_{\odot}$ model). 
$^{19}$F comes from the $^{14}$N($\alpha,\gamma$)$^{18}$F($\beta^+$)$^{18}$O($p,\alpha$)$^{15}$N($\alpha,\gamma$)$^{19}$F chain \citep{goriely89}. The protons mainly come from the $^{14}$N($n,p$)$^{14}$C reaction. The neutrons needed for the previous reactions are released by ($\alpha,n$) reactions, especially $^{13}$C($\alpha,n$) and $^{22}$Ne($\alpha,n$). Fig.~\ref{f19} shows the abundances of the species involved in the synthesis of $^{19}$F.
The additional $^{13}$C, $^{14}$N and $^{22}$Ne synthesized in rotating models largely contributes to boost the sequence described above and consequently the $^{19}$F production.

\begin{figure}[h!]
   \centering
      \includegraphics[scale=0.48]{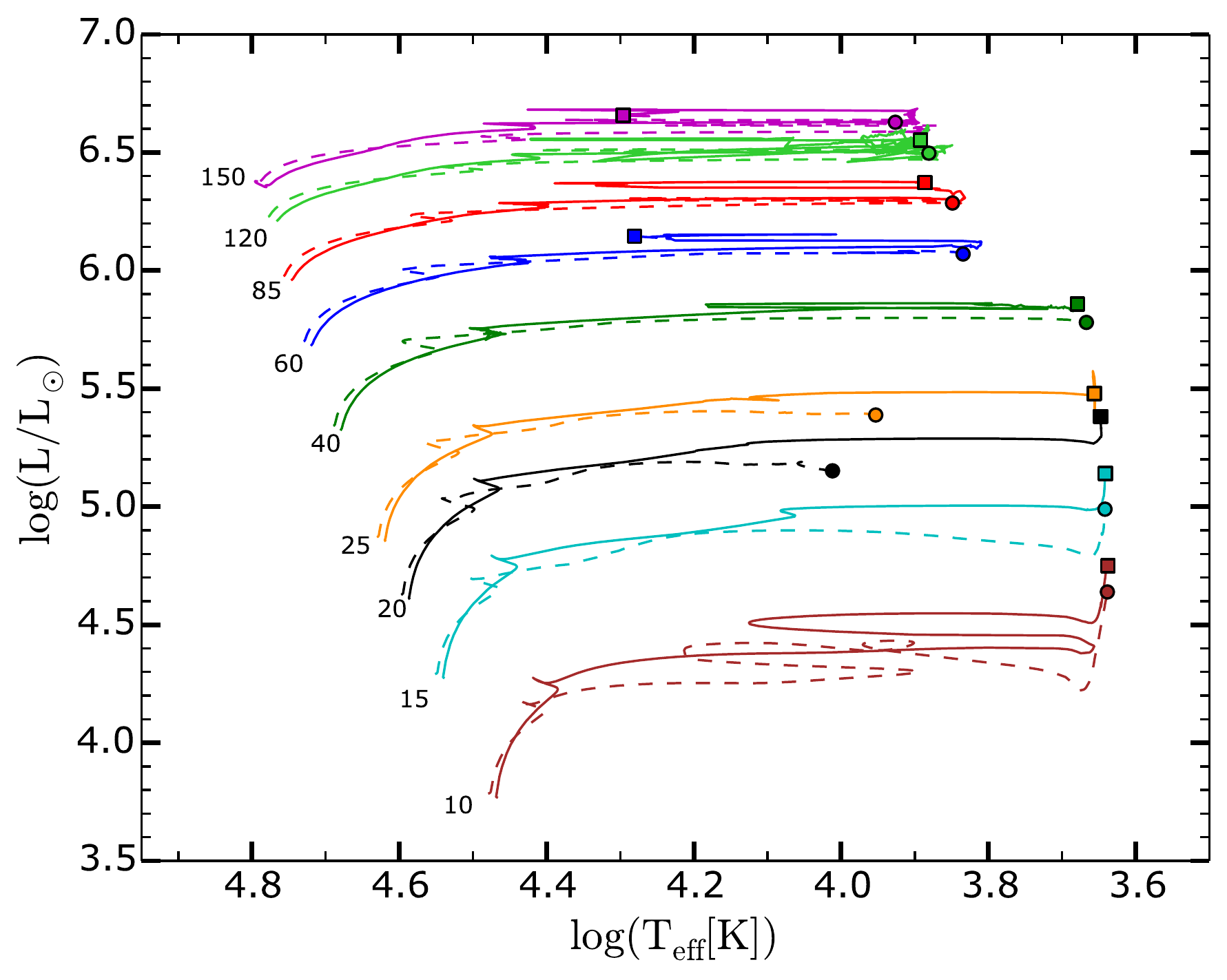}
   \caption{Tracks of the models in the Hertzsprung-Russell diagram. Dashed and solid lines show non-rotating and rotating models respectively. Circles and squares denote the endpoint of the evolution for the non-rotating and rotating models, respectively.}
\label{hrd}
    \end{figure}

In general, rotating models lose more mass during their evolution. It occurs mainly because of three reasons: 
\begin{itemize}
\item [$\bullet$] rotation increases the mass losses by line driven winds,
\item [$\bullet$] rotation changes the distribution of the chemical species in the stellar interior. It can modify the tracks in the HR diagram and therefore the mass loss experienced by the star,
\item [$\bullet$] rotation can also induce mechanical mass losses when the stellar surface reaches the critical velocity. 
\end{itemize}
The surface of the 25, 40 and 60 $M_{\odot}$ models reach the critical velocity at the end of the Main-Sequence so that mechanical mass-loss occurs. The mass lost due to that effect remains modest 
(less than 0.1 $M_{\odot}$). 
For the models of this work, the most important effect comes from the second reason mentioned above. 
After the Main-Sequence, rotating models have higher luminosities than non-rotating models (Fig.~\ref{hrd}). This is due to internal mixing, that tends to produce larger helium burning cores. The higher luminosity (1) increases directly the mass-loss rate and (2) makes the model more likely to enter the supra-Eddington regime. In this regime, additional mass loss occurs (c.f. Sect.~\ref{inputs}). The rotating 25 $M_{\odot}$ model becomes supra-Eddington close to the end of the core helium burning stage while its non-rotating counterpart never enters this regime. 
In the end, the rotating 25 $M_{\odot}$ model loses 8 more solar masses compared to the non-rotating model (Table~\ref{table:1}, last column).
Quickly after core He ignition, the rotating 60 $M_{\odot}$ model reaches $\log T_{\rm eff} \sim 3.8$ and  experiences a supra-Eddington stage that removes $\sim 8$ $M_{\odot}$. The stellar surface in then enriched in helium and makes the star going back to the blue (Fig.~\ref{hrd}). The non-rotating 60 $M_{\odot}$ enters the supra-Eddington regime only at the very end of core He-burning. Its surface is not much enriched in helium so that it stays red.

   \begin{figure*}
   \centering
      \includegraphics[scale=0.62]{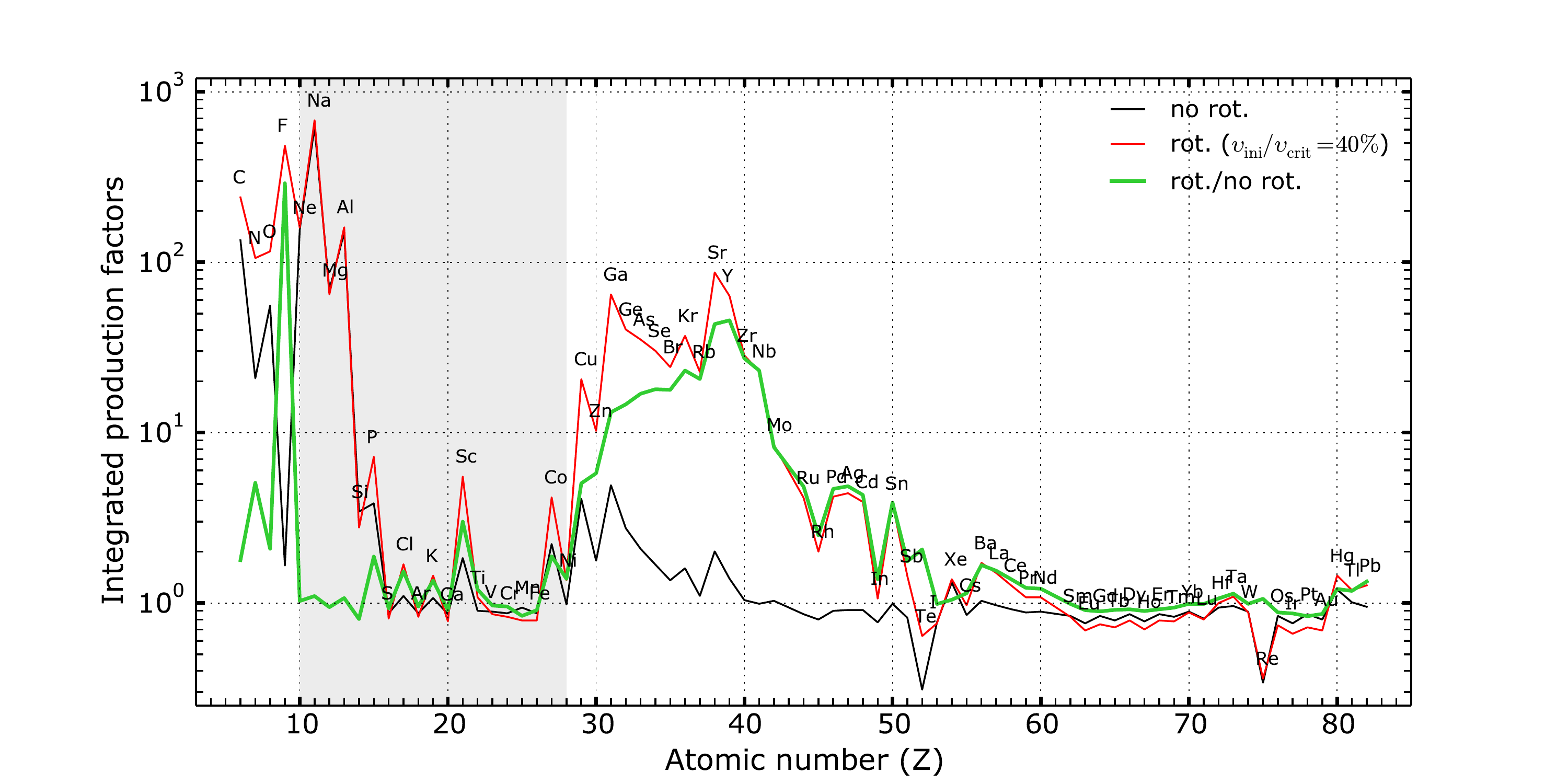}
   \caption{Integrated production factors $f_{i, \rm{int}}$ (Eq.~\ref{intpro}) for the population of non-rotating and rotating models. 
   The mass function of \cite{chabrier03} is used. The green line shows the ratio between the 2 curves. The mass cut is set using the relation of \cite{maeder92}. The grey area highlights the elements that are likely affected by explosive nucleosynthesis.}
\label{s0s4}
    \end{figure*}

\subsection{Integrated production factors}\label{intprod}
The integrated production factor $f_{i, \rm{int}}$ for an isotope $i$ is

\begin{equation}
f_{i, \rm{int}}  = \frac{ \int_{M_{\rm min}}^{M_{\rm max}} f_{i}(M_{\rm ini})~ \phi (M_{\rm ini}) ~\text{d} M_{\rm ini}}{\int_{M_{\rm min}}^{M_{\rm max}} \phi (M_{\rm ini})~ \text{d}M_{\rm ini} }
\label{intpro}
\end{equation}
where $M_{\rm min} = 10~M_{\odot}$, $M_{\rm max} = 150~M_{\odot}$, and $\phi (M_{\rm ini})$ is the initial mass function. Here we take the initial mass function of \cite{chabrier03} defined as $\phi (M_{\rm ini}) = AM_{\rm ini}^{-\alpha}$ with $A = 7.1~10^{-5}$ and $\alpha = 2.3$. The $f_{i, \rm{int}}$ factors were computed for the non-rotating and rotating (40 \% of critical velocity) models (black and red lines in Fig.~\ref{s0s4}). 

Because of the low weight associated with very massive stars (c.f. Eq.~\ref{intpro}), the contribution of such stars to the integrated pattern is small. The final pattern resembles the one of a $\sim 20~M_{\odot}$ model. The green line shows the ratio between the 2 factors. 
Strong differences occur between $30<Z<50$, especially around $Z=38$ (strontium). Also of interest is the fluorine, which is overproduced by more than 2 dex by the rotating population (c.f. Sect.~\ref{rotmodels}).

\begin{table}[h]
\scriptsize{
\caption{model label (column 1), initial mass (column 2), final mass (column 3), mass ejected through winds (column 4), mass of the H-free region (mass coordinate where the mass fraction of hydrogen $X(\rm ^{1}H)$ drops below 0.01, column 5), mass of the CO core (mass coordinate where the $^{4}$He mass fraction $X(\rm ^{4}He)$ drops below 0.01, column 6), remnant mass $M_{\rm rem}$ using the relation of \citet[][column 7]{maeder92}. \label{table:4}}
\begin{center}
\resizebox{8.7cm}{!} {
\begin{threeparttable}
\begin{tabular}{lccc|ccc} 
\hline 
\hline 
Model  		& $M_{\rm ini}$ & $M_{\rm fin}$ & $M_{\rm ej,wind}$ & $M_{\rm He}$ & $M_{\rm CO}$  &  $M_{\rm rem}$ \\ 
 & [$M_{\odot}$] & [$M_{\odot}$]     &	[$M_{\odot}$]      &	[$M_{\odot}$]	& [$M_{\odot}$] &	[$M_{\odot}$] 	\\
\hline 
10s0		&10	&9.93	&0.07&2.91   & 1.58   & 1.28	\\
10s4		&10	&9.78	&0.22&3.35   & 1.86    &1.36	\\
\hline
15s0		&15	&14.78	&0.22&4.67    &2.62    &1.56	\\
15s4		&15	&14.34	&0.66&5.74    &3.36    &1.75\\
\hline
20s0		&20	&19.87	&0.13&6.46    &4.04    &1.92\\
20s4		&20	&17.36	&2.64&8.21    &5.31    &2.24\\
\hline
25s0		&25	&24.68	&0.32&8.65    &5.88    &2.39\\
25s4		&25	&16.68	&8.32&10.85   &7.53    &2.80\\
25s7		&25	&16.21	&8.79&10.91   &7.56    &2.81\\
25s7B$^{\rm a}$&25&16.01	&8.99&10.96   &7.62  &  2.82\\
\hline  
40s0		&40	&34.10   &5.90&15.29   &11.75   &3.80	\\
40s4		&40	&24.60   &15.40&19.10   &14.66   &4.54	\\
\hline  
60s0		&60	&44.17   &15.83&25.36   &20.94   &6.44	\\
60s4		&60	&40.81   &19.19&30.77   &25.35   &7.77	\\
 \hline
85s0		&85	&59.33   &25.67&37.86   &32.86   &9.90	\\
85s4		&85	&58.27   &26.73&45.57   &39.30   &11.73	\\
\hline
120s0		&120	&82.45   &37.55&54.30  & 50.01  & 14.69	\\
120s0B$^{\rm a}$	&120	&83.18   &36.82&54.46   &49.35   &14.51	\\
120s4		&120	&85.81   &34.19&65.28   &61.78  & 17.95\\
\hline
150s0		&150	&100.26  &49.74&70.77   &65.34   &18.94	\\
150s4		&150	&99.59   &50.41&83.83   &80.04   &23.04	\\
\hline
\end{tabular}
\begin{tablenotes}
\item[a] Models computed with the rate of $^{17}$O($\alpha,\gamma$) divided by 10.
\end{tablenotes}
\end{threeparttable}
}
\end{center}
}
\end{table}

\subsection{Faster rotation}

Increasing the initial rotation rate from 40\% to 70\% of the critical velocity for the $25$ $M_{\odot}$ model allows the production of s-elements up to $Z \sim 60$ (dashed purple line in Fig.~\ref{zm3rot}). Compared to the 25 $M_{\odot}$ model with slower rotation, Sr and Ba are overproduced by $\sim$ 0.2 and 1 dex respectively. This is shown by the small blue and red arrows in Fig.~\ref{yields}. Fast rotation boosts more the second than the first s-process peak with respect to the 40\% case. This is because faster rotation gives more $^{22}$Ne, hence more neutrons and a higher source (neutrons) over seed (heavy elements) shifts the production of s-elements towards higher masses \citep{gallino98}.

\subsection{Lower $^{17}$O($\alpha,\gamma$) rate}

When the $^{17}$O($\alpha,\gamma$) rate is reduced in the fast rotating 25 $M_{\odot}$ model, the source over seed ratio is also increased since more neutrons are recycled. This allows the production of even more massive elements, up to Pb (dotted line in Fig.~\ref{zm3rot}). The largest difference between the two fast rotating models with different reaction rates occurs for $Z>55$. In particular, Ba and Hg are overproduced by more than 1 dex. Reducing the rate of $^{17}$O($\alpha,\gamma$) in the non-rotating 120 $M_{\odot}$ model boosts the production of light s-elements by $\sim 0.5$ dex but does not allow the significant production of elements heavier than $Z \sim 50$. A better knowledge of the $^{17}$O($\alpha,\gamma$) rate is crucial to better constrain the production of the s-elements, especially from the second peak, which are largely affected when changing this nuclear rate.

   \begin{figure*}
   \centering
   \begin{minipage}[c]{.49\linewidth}
   \centering
      \includegraphics[scale=0.5]{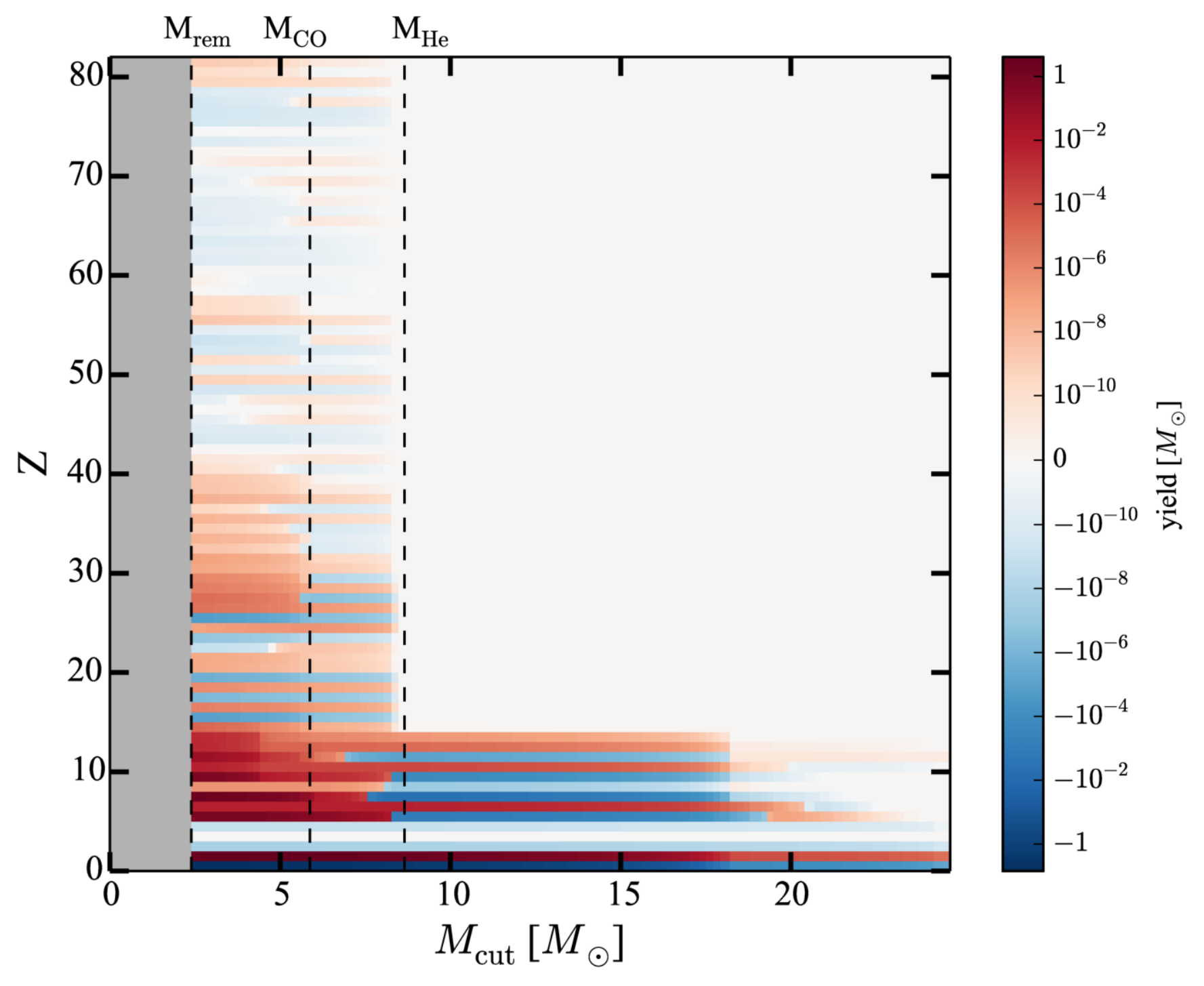}
   \centering
   \end{minipage}
   \begin{minipage}[c]{.49\linewidth}
   \centering
      \includegraphics[scale=0.5]{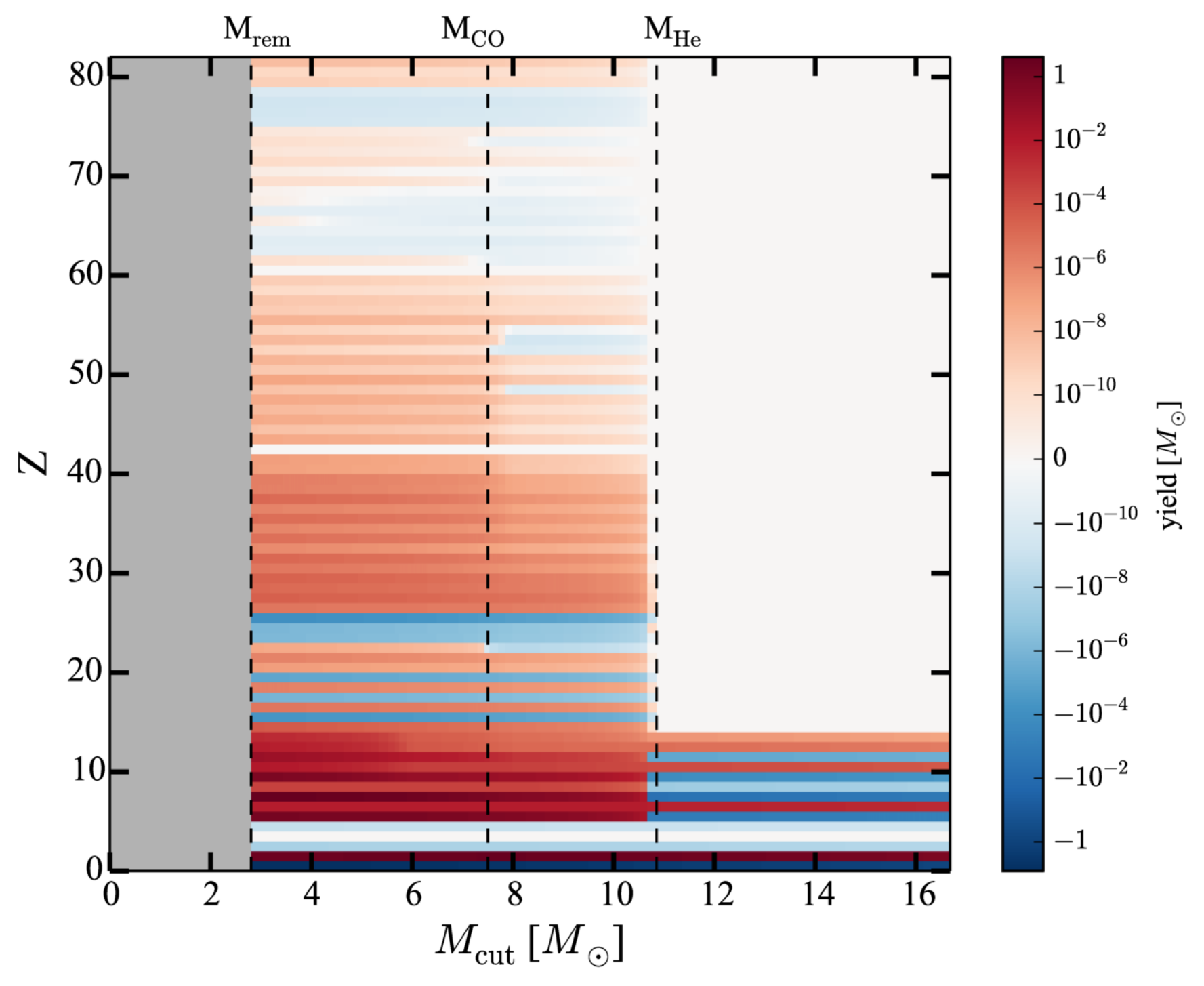}
   \centering
   \end{minipage}
   \begin{minipage}[c]{.49\linewidth}
   \centering
      \includegraphics[scale=0.5]{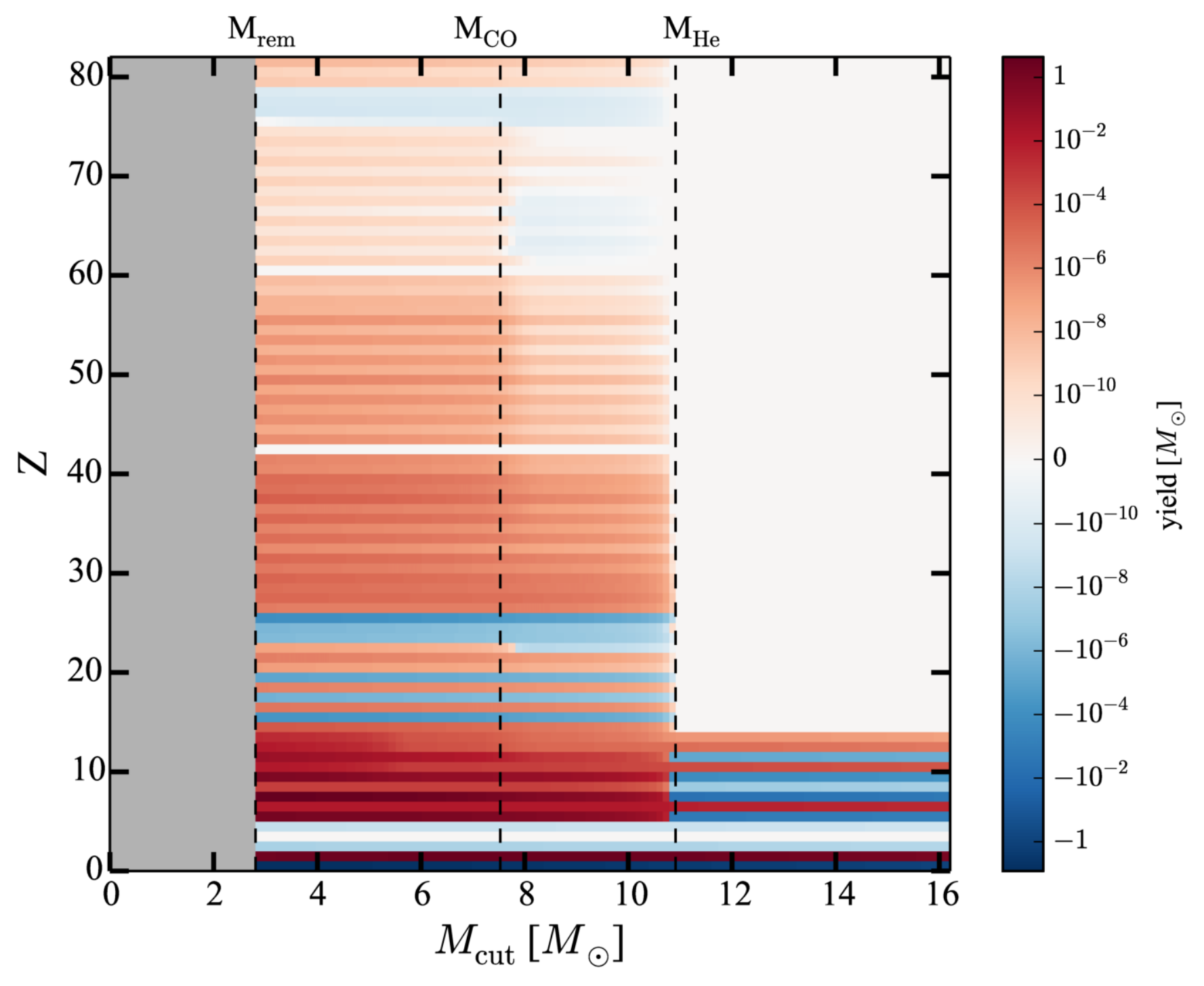}
      \centering
   \end{minipage}
   \begin{minipage}[c]{.49\linewidth}
   \centering
      \includegraphics[scale=0.5]{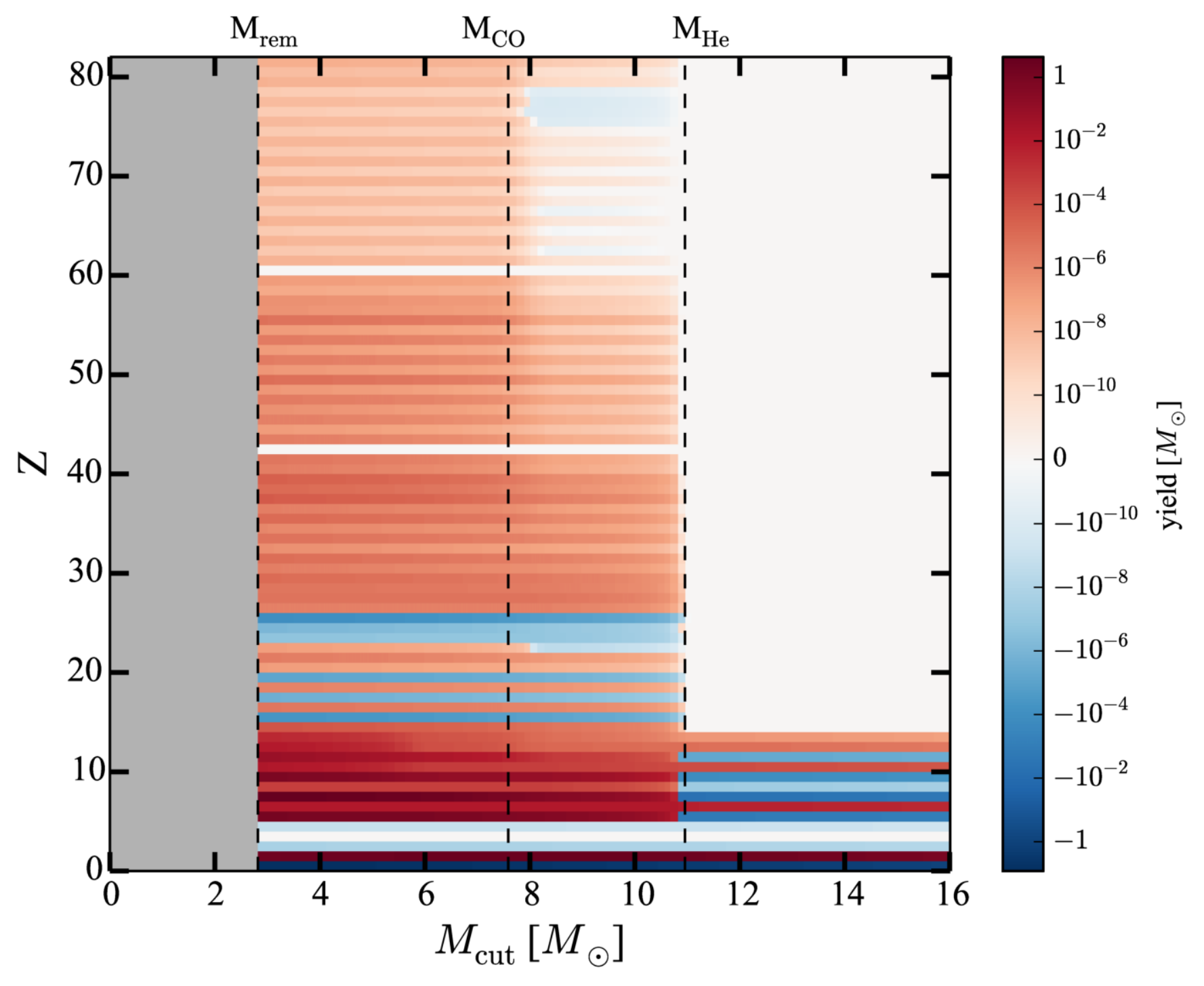}
      \centering
   \end{minipage}
   \caption{
   Yields of elements (characterized here by the atomic number Z) for different values of the mass cut.
   The color map shows the yields for the non-rotating 25 $M_{\odot}$ (top left panel), rotating 25 $M_{\odot}$ (top right), fast rotating 25 $M_{\odot}$ (bottom left) and fast rotating 25 $M_{\odot}$ with lower $^{17}$O($\alpha,\gamma$) (bottom right). The ticks labelled $M_{\rm rem}$ show the location of the remnant mass using the relation of \cite{maeder92} (last column of Table~\ref{table:4}). 'CO' and 'He' denote the location of the top of the CO and He core respectively (fifth and sixth columns of Table~\ref{table:4}).}
\label{map25}
    \end{figure*}

   \begin{figure*}
   \centering
   \begin{minipage}[c]{.49\linewidth}
   \centering
      \includegraphics[scale=0.5]{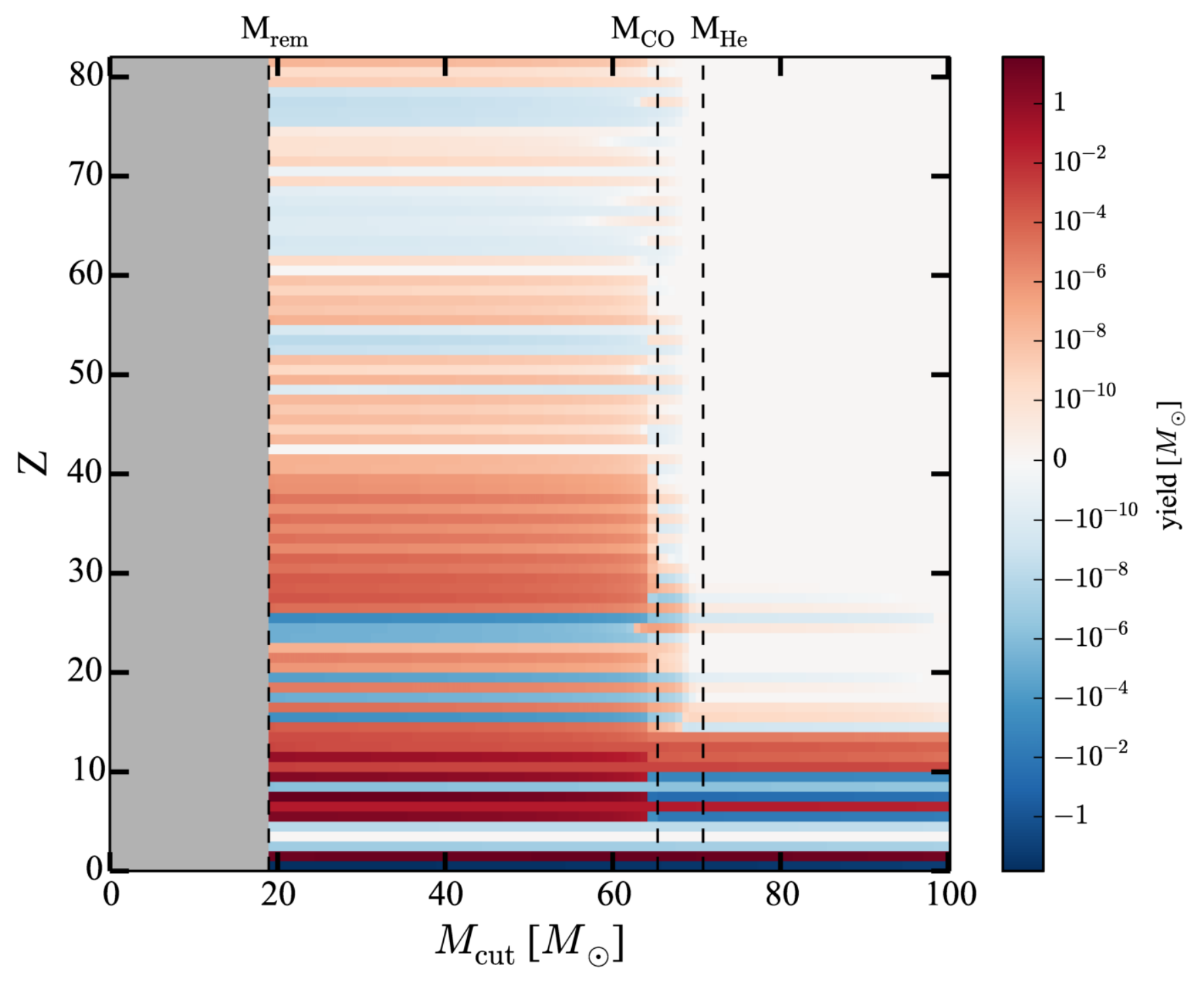}
   \centering
   \end{minipage}
   \begin{minipage}[c]{.49\linewidth}
   \centering
      \includegraphics[scale=0.5]{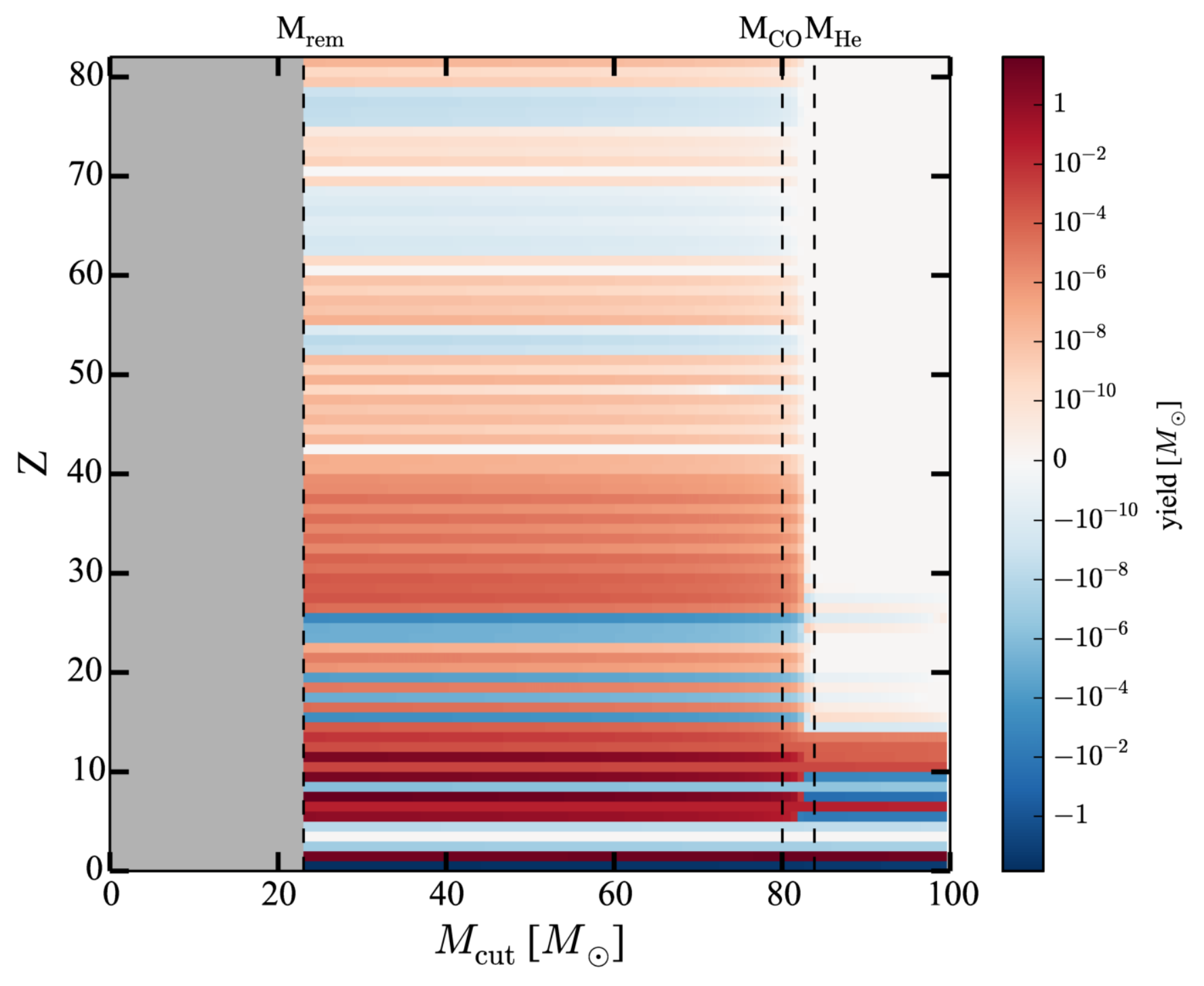}
   \centering
   \end{minipage}
   \caption{Same as Fig.~\ref{map25} but for the non-rotating 150 $M_{\odot}$ (left) and rotating 150 $M_{\odot}$ (right).}
\label{map150}
    \end{figure*}

\section{Effect of the mass cut and yield tables}\label{effmcut}

\subsection{Effect of the mass cut}\label{subeffmcut}

In the previous section, we discussed the yields assuming a specific mass cut \citep[following][]{maeder92}. However, how massive stars explode is still poorly constrained, and even less if rotation is included. 
It is generally difficult to strongly state which part of the star is expelled and contributes to the chemical enrichment of the interstellar medium (ISM). 
In what follows, we discuss the effect of varying the mass cut. In the yield tables provided with this work, the mass cut is let as a free parameter. 

Figures \ref{map25} and \ref{map150} show the dependence of the yields on the mass cut for the 25 and 150 $M_{\odot}$ models. 
They show how elements are produced (positive yield, red color) or destroyed (negative yield, blue color) when varying the mass cut between the final mass $M_{\rm fin}$ and the remnant mass $M_{\rm rem}$ of the model. 
The gaps at $Z=43$ and $61$ in every panel correspond to the elements Tc and Pm, which have no stable isotope and are consequently neither produced nor destroyed in the final yields. 
Considering the fast rotating 25 $M_{\odot}$ model with lower $^{17}$O($\alpha,\gamma$) (Fig.~\ref{map25}, bottom right panel), we see that a mass cut below $\sim 10.5$ $M_{\odot}$ is needed to expel s-elements with $27<Z<60$ and a mass cut below $\sim 7.5$ $M_{\odot}$ (corresponding to the bottom of the He-burning shell) to expel s-elements with $Z>60$.
Interestingly, elements with $60<Z<70$ (among them Eu) are exclusively produced in the inner regions of the two fast rotating 25 $M_{\odot}$ models (bottom panels of Fig.~\ref{map25}). Also, out of these two models, only the one with a lower $^{17}$O($\alpha,\gamma$) is able to produce elements with $75<Z<80$.

Models including rotation generally lose more mass during their evolution (c.f. Sect.~\ref{rotmodels}) and have larger helium cores so that s-elements are located closer to the stellar surface at the end of the evolution. A large mass cut (i.e. close to the surface) will then already eject some s-elements in the case of the rotating models. Without rotation, s-elements are located deeper inside the star so that a smaller mass cut (i.e. deeper inside the star) is required to eject these elements. As an example, $M_{\rm cut} = 70$ $M_{\odot}$ will eject some s-elements for the rotating 150 $M_{\odot}$ model while it will not for the non-rotating 150 $M_{\odot}$ model (Fig.~\ref{map150}).
Ejecting deeper layers likely requires a more powerful explosion. Consequently, rotation in massive stars not only boosts the production of s-elements but might also make it easier to expel these elements. This could be viewed as a possible indirect effect of the rotation on providing more s-elements to the ISM.

\subsection{Yield tables}\label{yieldst}

Electronic tables of yields are available on the web\footnote{\href{https://www.unige.ch/sciences/astro/evolution/en/database/}{https://www.unige.ch/sciences/astro/evolution/en/database/} 
}. 
In these tables, the mass cut $M_{\rm cut}$ is varied between the final mass of the considered stellar model and the remnant mass $M_{\rm rem}$ from \cite{maeder92}. 100 values of $M_{\rm cut}$ are considered for each models, equally spaced between $M_{\rm rem}$ and $M_{\rm fin}$. The entire table therefore contains 21 stellar models times 100 $M_{\rm cut}$ that means 2100 different ejecta compositions.
A part of the yield table for the rotating 25 $M_{\odot}$ is shown in Table \ref{table:5}. The first value of $M_{\rm cut}$ (column 2) is equal to the final mass of the model $M_{\rm fin}$ (given in Table~\ref{table:4}). It corresponds to the case where only the mass loss through stellar wind is taken into account (also given in Table~\ref{table:4}). 
In all the other cases, the yield of an isotope is the sum of the yields in the wind plus the yields in the material ejected by a supernova of the indicated mass cut. 
The last value of $M_{\rm cut}$ (last column in Table \ref{table:5}) corresponds to the case where all the material above $M_{\rm cut} = M_{\rm rem}$ is ejected (also the stellar wind is taken into account). 
Yields below $10^{-15}$ $M_{\odot}$ in absolute value are set to zero. 
Fig.~\ref{map25} and \ref{map150} are graphic representations of such tables.

\section{Summary and discussions}\label{concl}

We computed a grid of 21 models with and without rotation, at $Z=10^{-3}$ and with initial masses between 10 and 150 $M_{\odot}$. Rotating models were computed with an initial rotation of 40 \% of the critical velocity. One model was computed with 70 \% of the critical velocity and 2 models with the rate of $^{17}$O($\alpha,\gamma$) divided by 10. With this paper, we provide tables of yields including the effect of varying the mass cut. 

The main result of this work is that rotation has the strongest impact on s-element production for $20< M_{\rm ini} <60$ $M_{\odot}$. The first s-process peak is the most affected by rotation. In the 25 $M_{\odot}$ rotating model, the yield of $^{88}$Sr is increased by $\sim 3$ dex (Fig.~\ref{yields}). Although to a smaller extent, the second and third peak are also affected: $^{138}$Ba and $^{208}$Pb are overproduced by $\sim 1$ dex.
Faster rotation boosts even more the s-element production in the range $40<Z<60$.
Taking a reasonably lower $^{17}$O($\alpha,\gamma$) reaction rate in the fast rotating model overproduces the s-elements with $Z>55$ (among them Pb) by about 1 dex compared to the standard fast rotating model.

\begin{table*}[t]
\scriptsize{
\caption{Yields in $M_{\odot}$ (Eq.~\ref{yie}) of the rotating 25 $M_{\odot}$ model. Only a part of the table is shown. The full table (including all models, mass cuts and isotopes) is available with the online version of the paper. Each column corresponds to a given mass cut in $M_{\odot}$. The first mass cut (16.68 $M_{\odot}$) corresponds to the final mass of the model. In that case, the yields are only from stellar winds. The last mass cut corresponds to $M_{\rm rem}$ (c.f. Table~\ref{table:4}). In between these two extreme values, 100 equally spaced $M_{\rm cut}$ are taken. \label{table:5}}
\begin{center}
\resizebox{18.0cm}{!} {
\begin{tabular}{c|cccccccccc}
\hline
& 16.68 & 16.54 & 16.40 & 16.26 & ... & ... & 3.22 & 3.08 & 2.94 & 2.80 \\
\hline
 $^{1}$H    & $-7.113$($-01$) & $-7.113$($-01$) & $-7.113$($-01$) & $-8.189$($-01$) & ...& ... & $-8.018$($+00$) & $-8.123$($+00$) & $-8.228$($+00$) & $-8.333$($+00$) \\
 $^{2}$H    & $-1.209$($-04$) & $-1.209$($-04$) & $-1.209$($-04$) & $-1.270$($-04$) & ... & ...& $-3.166$($-04$) & $-3.187$($-04$) & $-3.207$($-04$) & $-3.227$($-04$) \\
 $^{3}$He    & $-2.434$($-04$) & $-2.434$($-04$) & $-2.434$($-04$) & $-2.587$($-04$) & ... & ...& $-7.742$($-04$) & $-7.800$($-04$) & $-7.858$($-04$) & $-7.916$($-04$) \\
 ...&...&...&...&...&...&...&...&...&...&... \\
 ...&...&...&...&...&...&...&...&...&...&... \\
  $^{207}$Pb    & $0$ & $0$ & $0$ & $0$ & ...& ... & $+1.124$($-09$) & $+1.149$($-09$) & $+1.173$($-09$) & $+1.198$($-09$) \\
 $^{208}$Pb    & $0$ & $0$ & $0$ & $0$ & ...& ... & $+3.922$($-09$) & $+4.024$($-09$) & $+4.125$($-09$) & $+4.226$($-09$) \\
 $^{209}$Bi    & $0$ & $0$ & $0$ & $0$ & ... & ...& $+1.384$($-10$) & $+1.423$($-10$) & $+1.462$($-10$) & $+1.502$($-10$) \\
\hline
\end{tabular}
}
\end{center}
}
\end{table*}

\subsection{Initial rotation of the models}

The boost of the s-process element production in massive stars is obtained here through rotational mixing. The importance of the boost depends among other parameters on the initial angular momentum content of the star, here determined by the choice of the surface rotation velocity on the ZAMS where the star is supposed to rotate as a solid body. The present results have been obtained only for one initial rotation for each initial mass and of course, to obtain a broader view of the impact of rotation, families of models with different initial rotation rates should be computed for each initial masses. Here, to limit the computational time (that is significant when 
following the changes in the abundances of such a large number of isotopes), we focused on a particular choice (40\% the critical velocity at the ZAMS). We adopted this value for the following reasons: first, at solar metallicity, this choice is consistent with the peak of the velocity distribution of young main-sequence B-type stars (c.f. Sect.~\ref{inputs}).
Second, we wanted to use the same initial rotations than those used in F12 and F16 in order to check how some changes brought to the code since these computations may have affected the results. Since there are no observational constraints about the velocity distributions at the metallicity considered here, it is difficult to know whether such a choice is  representative or not. At the moment, in absence of such confirmation, we can see the present computations as an exploration on how the boost of the s-process due to rotation varies as a function of the initial masses over a large range of initial masses. The reader has to keep in mind that the absolute values of the yields depends here on the choice of the initial rotation.

\subsection{Model uncertainties}\label{moduncer}

One has also to keep in mind that the yields of stellar models are affected by several sources of uncertainty. By changing the rate of $^{17}$O($\alpha, \gamma$), we provided an example on how current nuclear rate uncertainties can affect the yields. The three other key reactions for s-process in massive stars (shown in Fig.~\ref{figcomprates}), that are still not completely constrained, add another source of uncertainty in the yields. Uncertainties on neutron-captures and $\beta$-decay rates also affect the s-process yields by a factor of 2 at maximum, in general \citep{nishimura17, nishimura18b}.

Also, even if we know that stars rotate, important uncertainties remain on the effects of rotation in the stellar interiors, hence on the s-process yields of rotating stars. 
The production of s-elements is highly sensitive to the amount of $^{22}$Ne available, which in turn, depends on stellar evolution inputs such as the way rotational mixing (also convection) is treated in the code. Different recipes exist in the literature for the horizontal diffusion \citep{zahn92, maeder03, mathis04} and the shear diffusion coefficients \citep{talon97, maeder97, maeder13} that govern the transport of chemical elements \citep[see][for a review]{meynet13}. Different combinations of these coefficients will lead to a higher/smaller production of extra $^{22}$Ne, hence a possibly different production of s-elements. 
Such uncertainties might be at the origin of the differences between this work and the recent work of \cite{prantzos18}. They used a chemical evolution model to discuss the abundance evolution of elements up to uranium in the Milky Way. They included yields of rotating massive stars from Chieffi \& Limongi. 
Elements from Ba to Pb are generally overproduced compared to our models.
Since these stellar models are not published yet, we do not know the detailed physics ingredients and cannot do extensive comparisons.

\subsection{Fluorine and s-elements}

From discussions in Sect.~\ref{intprod} and Fig.~\ref{s0s4}, we note that the production of s-element in massive rotating stars should be correlated with the production of several light elements, particularly fluorine. 
This correlation might be found in the next generation of low mass halo field stars. 
Importantly, AGB stars are also believed to contribute to the production of both fluorine and s-elements \citep[e.g.][]{jorissen92, lugaro04, lugaro08, abia10, karakas10, bisterzo10, gallino10}, leaving open the possibility for AGB and massive stars to be responsible for such abundance patterns. 
In addition to AGB and massive rotators, the $\nu$-process in core-collapse supernovae is also generally expected to contribute to fluorine production \citep{woosley77,woosley90,kobayashi11,izutani12}.
Recent studies suggest however that both the $\nu$-process in supernovae \citep{jonsson17} and AGB stars \citep{abia15} might be insufficient to explain the observed evolution of fluorine in the solar neighborhood.
This potentially makes rotating massive stars interesting complementary fluorine sources that might improve the agreement between galactic chemical evolution models and observations \citep{meynet00, palacios05}.

Around the metallicity considered in this work ([Fe/H] $\sim -2$), the few iron-poor halo field stars whose fluorine abundance was determined are generally F-rich \citep[][]{otsuka08, lucatello11, li13, schuler07}. One star (HD5223) with [Fe/H] $\sim -2$ has both fluorine and heavy element abundances available. It is enriched in F, Sr, Ba and Pb \citep{goswami06,lucatello11} and shows radial velocity variations \citep{mcclure90}. The enhancement in Pb may not be reproduced by the massive stellar models of the present work. 
A mass transfer episode from an AGB stars companion may be the main process for explaining the abundances of HD5223. 
Further determinations of fluorine and s-elements abundances in metal-poor stars should help testing  stellar model predictions.

\subsection{Rotation from solar to very low metallicity}

As a final note, we would like to emphasize here that the impact of rotation on the evolution of stars (in particular on the stellar yields) allows to unify in a same theoretical framework the properties of stars observed, for instance, in the solar neighbourhood to the properties of stars and their impact on nucleosynthesis at very low metallicities \citep[see e.g.][]{maeder15a, chiappini13}.

Let us first recall that rotational mixing has been first included in stellar models to account for surface enrichments observed at the surface of main-sequence B-type stars in the solar neighbourhood \citep[see e.g.][and references therein]{maeder12}. In general, rotating models need to be calibrated in order to constrain the efficiency of rotation-induced mixing (c.f. Sect.~\ref{inputs}). 
In the present work, the value of $f_{\rm energ}$ (Eq.~\ref{dshtz97}) is chosen in order for solar metallicity models with initial masses around 15 $M_{\odot}$ to fit the the averaged observed chemical enrichments of Galactic B-type stars rotating with an average surface velocity.
Although the calibration can be done using different observations \citep[][for instance, used a sample of B-type stars in the Large Magellanic Cloud]{brott11}, at very low metallicities, there are no observation allowing to check whether a different value of $f_{\rm energ}$ would be needed. At the moment the most reasonable choice is to keep this quantity constant. 
Once the calibration is done, the physics describing the transport processes of both chemical elements and angular momentum due to rotation is not changed. 
As a consequence, the results of the stellar models for other initial masses and metallicities can be seen as stellar models predictions. 
Interestingly, when this physics is used for low metallicity rotating stars of both intermediate and high masses, the rotational mixing produces, \textit{without any artificial tuning}, primary nitrogen production \citep{meynet02b}. 
Rotating massive star models have been invoked to explain the N/O plateau shown by metal-poor halo stars, the C/O upturn \citep{chiappini06} and provided predictions concerning the $^{12}$C/$^{13}$C ratio \citep{chiappini08}.

Rotation is also interesting to explain the CEMP stars with [Fe/H] $<-3$ that are not highly enriched in s- and/or r-elements \citep{meynet06, hirschi07, meynet10, joggerst10, takahashi14, maeder15a, maeder15b, choplin16, choplin17a}. It was reported by \cite{placco14c} that 43 \% of stars with [Fe/H] $<-3$ are CEMP (with [C/Fe] $>0.7$ and excluding the stars showing clear overabundances of neutron-capture elements). A major difference between CEMP stars and normal metal-poor halo stars is likely due to the degree of mixing of the cloud of interstellar material from which these two types of stars formed. CEMP stars likely formed from pockets of ISM that have been enriched by the ejecta of a few objects, may be only one, while normal halo stars are likely formed from a much better mixed reservoir in which the ejecta of many more sources have accumulated. In both cases (normal halo stars and at least some CEMP stars), rotational mixing provides a very interesting mechanism for explaining the surface abundances of many of these objects, while still being able to account for observed features of massive stars at solar metallicity. Of course alternative explanations exist \citep[e.g.][]{umeda03,limongi03, iwamoto05, tominaga14, clarkson18} and in the future some specific signatures will hopefully allow to decide which of these models or combination of models are the most probable. 
A comparison between the different models for explaining the most iron-poor stars is beyond the scope of this work, which focuses on higher metallicities. 
We plan to compute similar models as done in the present paper but with a lower metallicity ([Fe/H] $\sim -4$).

\begin{acknowledgements} 
This work was supported by the SNF project number 200020-172505. 
Authors acknowledge support from the ''ChETEC'' COST Action (CA16117), supported by COST (European Cooperation in Science and Technology).
RH acknowledges the European Research Council under the European Union's Seventh Framework Programme (FP/2007-2013)/ERC Grant Agreement no. 306901. 
RH acknowledges support from the World Premier International Research Center Initiative (WPI Initiative), MEXT, Japan. CC acknowledges support from DFG Grant CH1188/2-1.
\end{acknowledgements}

\bibliographystyle{aa}
\bibliography{biblio.bib}

\end{document}